\documentclass[10pt,tightenlines,eqsecnum,floats,aps,nofootinbib,prd]{revtex4-2}
\usepackage{amsmath,amssymb,amsfonts,amsthm,amscd}
\usepackage{graphicx}
\usepackage{amsmath}
\usepackage{amssymb}
\usepackage{graphicx}
\usepackage{cancel}
\usepackage{slashed}
\usepackage{mathtools}
\usepackage{tcolorbox}
\usepackage{tikz-cd}
\usepackage{tikz}
\usepackage{mathrsfs}
\usepackage{relsize}
\usepackage{natbib}
\usepackage[margin=1in]{geometry}
\usepackage{hyperref}
\usepackage[normalem]{ulem}

\usetikzlibrary{shapes.geometric, arrows}
\tikzstyle{startstop} = [rectangle, rounded corners, minimum width=3cm, minimum height=1cm,text centered, draw=black, fill=red!30]
\tikzstyle{arrow} = [thick,->,>=stealth]
\begin{document}
\title{Classical General Relativity as a Non-Conservative Action-Dependent Field Theory}
\author{Callum Bell}
\email{c.bell8@lancaster.ac.uk}
\affiliation{Department of Physics, Lancaster University, Lancaster UK}

\author{David Sloan}
\email{d.sloan@lancaster.ac.uk}
\affiliation{Department of Physics, Lancaster University, Lancaster UK}

\begin{abstract}
    Previous work has provided the mathematical framework within which to analyse dynamical similarities for classical theories of fields. This formalism has been extended to those theories which, in addition to scaling symmetries, also possess gauge degrees of freedom. In this article, we apply these ideas to the analysis of the first-order Palatini formulation of General Relativity. It is shown that the conformal mode of the spacetime metric may be identified as the generator of a dynamical similarity. Further, we demonstrate that the dynamical content of the Hilbert-Palatini action may be reformulated in terms of an action-dependent field theory, which makes no reference to the conformal mode. Finally, we consider the linearised limit of the equations of motion derived from the scale-reduced action. We find that, in the harmonic gauge, the first-order metric perturbations satisfy a free wave equation, as expected. However, the elimination of the conformal factor requires a qualitative reinterpretation of the physics at second order. Conventionally, one considers the second-order perturbations to be sourced by quadratic combinations of first-order terms. These are packaged into an object identified as an `effective stress-energy tensor'. This interpretation must be amended for the action-dependent theory, where the presence of terms that couple the action sector with the geometrical degrees of freedom shows that our construction is inherently non-conservative.
\end{abstract}
\maketitle
\section{Introduction}\label{Sec:Introduction}
As a classical field theory, General Relativity (GR) may be formulated in a variety of equivalent ways, with the particular choice of formalism guided by the structural aspects of the theory one wishes to highlight. Of interest to us in the present context are those formulations which isolate the conformal mode - a degree of freedom that may be considered to represent a local choice of `metre stick' \cite{brown2005conformal}. In this framework, it is natural to express the full spacetime metric as a product of a scalar degree of freedom (encoded by the conformal mode) and a symmetric, rank-two covariant tensor of fixed determinant. Such a factorisation provides a clean separation of the metric degrees of freedom into something we might schematically describe as `local scale $\times$ conformal geometry'.\\

Motivated by the elegance and computational power of symplectic geometry in the study of mechanical systems, there has been growing interest in alternative geometrical frameworks, particularly those which incorporate non-conservative effects, that cannot be captured by terms in a conventional Lagrangian. The most relevant class of examples for us are those based on a so-called Herglotz variational principle \cite{cannarsa2019herglotz}, where the action itself is incorporated into the space of dynamical variables. Such theories have recently been explored in gravitational settings \cite{gaset2022variational,lazo2017action}; however, the nature of the action dependence of these models is relatively simplistic, and somewhat lacks physical motivation.\\

In this work, we show that, from the perspective of the evolution of the dynamical observables, the conformal mode is a redundant degree of freedom. This is done by explicitly constructing an action which reproduces the dynamical content of the Hilbert-Palatini action, but does so without reference to the conformal mode. Our description exhibits both a reduced gauge symmetry, where the symmetry group is restricted to those diffeomorphisms which are volume-preserving, and also a non-conservative structure. This latter property is made manifest through a Lagrangian which is explicitly action-dependent. Our objective is to present the construction of this action-dependent theory, and to analyse its physical implications. This is first carried out generically, at the level of the field theory, before considering a simple discussion of gravitational wave dynamics.\\

Our work is structured as follows: we begin with a brief motivational outline, in which we discuss the notion of superfluous mathematical structure, and how the introduction of such should be avoided when constructing mathematical models of the natural world. In section (\ref{Sec:ReductionofMultisymplecticFieldTheory}), we substantiate these ideas in a more mathematical setting, describing how the hallmark of a theory possessing redundant mathematical structure is the presence of a scaling symmetry, known as a \textit{dynamical similarity} \cite{sloan2018dynamical}.\\

In order to establish notational conventions, sections (\ref{Sec:OrthonormalFrames}) and (\ref{Sec:GravActions}) contain a very brief discussion of the first-order Palatini formulation of GR. The main results of our work, in which we identify a dynamical similarity, and construct an action-dependent Lagrangian, which reproduces the Hilbert-Palatini dynamics, is the subject of section (\ref{Sec:FirstOrderFrictionalAction}). In the interest of presenting a simple discussion of gravitational wave dynamics, it will be of benefit to recast our results in the second-order formalism, expressing the Lagrangian in terms of the metric, its derivatives, and the action variables. This procedure is carried out in section (\ref{Sec:Metric}), after which we compute the field equations. Upon linearising these field equations around a flat background, we find that the physical dynamics of the first-order perturbations are unaffected by the action-dependent nature of our theory. The sole effect of the dissipative sector is to reorganise the gauge degrees of freedom.\\

Expanding the equations to second order, the conventional interpretation is that the self-interaction of gravitational radiation may be described in terms of an effective conserved energy-momentum tensor. Such an interpretation fails for the action-dependent theory, where terms that couple the metric perturbations and the action variables must be understood in an intrinsically non-conservative context. We emphasise that this, along with all other `differences' between our construction and standard GR, are entirely interpretational. Mathematically speaking, the dynamical content of the two formulations is equivalent.\\

The results of this paper constitute the first steps towards a complete Hamiltonian analysis of GR as an action-dependent field theory. We shall assume throughout that our spacetime manifold $M$ is closed and compact, allowing us to present the reduced theory, without the hindrance of additional complications arising from boundary terms. Of course, such models do not satisfy the chronology condition, and so we identify as future work a more complete framework, applicable to physical spacetimes. We conclude our work with a brief summary of the results obtained, together with a number of additional lines of further investigation.
\section{Leibniz's Principle: The Identity of Indiscernibles}\label{Sec:Leibniz}
Whilst philosophical argumentation is frequently employed in conjunction with physical insight, there are notable instances in which philosophical principles have played a decisive role in the deduction of natural law. In the current context, we refer to a principle due to Leibniz, known as the Identity of Indiscernibles \cite{leibniz1989discourse}. This posits that if one has a theory that assigns distinct ontological status to two situations that are empirically indistinguishable, then the theory contains surplus structure, and the distinction should not be regarded as physically meaningful.\\

A familiar example is provided by uniform gravitational fields and uniform acceleration, where locally, there is no experiment that can be performed to distinguish one from the other. From the Leibnizian perspective, it is therefore essential that a theory of the gravitational interaction should not assign distinct ontological status to these two situations. Of course, these ideas are not unfamiliar: The Equivalence Principle, one of the foundational ideas of GR, embodies precisely this philosophy, by asserting that gravitational effects associated with a uniform field may be locally removed by an appropriate choice of frame. Consequently, only quantities invariant under changes of description should be regarded as physically meaningful. In this work, we argue that this principle has not yet been implemented to its fullest extent: the theory still distinguishes between certain configurations that are empirically indistinguishable, suggesting the existence of an additional redundancy. As we shall discuss extensively in due course, the remaining redundancy should be understood as a symmetry between configurations differing by a notion of scale.\\

These ideas provide the motivation for our study of dynamical similarities. Mathematically, the dynamics of a classical system may be deduced either in terms of a variational principle, or within the framework of symplectic geometry. While closely related, it is productive to consider each of these perspectives separately. From the variational point of view, one frequently encounters certain combinations of rescalings of the canonical variables (e.g $x\rightarrow \lambda^2 x$, $\dot{y}\rightarrow \lambda^{-1} \dot{y}$, for $\lambda\in\mathbb{R}$) that transform the Lagrange density only by an overall multiplicative factor ($\mathcal{L}\rightarrow \lambda^{\Lambda}\mathcal{L}$). Such a rescaled object shares its extremal points with the original Lagrange density, and so describes the same classical dynamics. From the symplectic viewpoint, these transformations are active, non-trivial maps on the system's phase space, rather than reparameterisations or simple changes of coordinates. They generate distinct trajectories, that nevertheless induce identical observable dynamics.\\

Following the principle of the Identity of Indiscernibles, the redundancy induced by a scaling symmetry is considered surplus structure. As described below, such structure may be eliminated by forming the quotient of the system's phase space by the (one-dimensional) orbits of the symmetry generator. The result is a description in which each observable configuration evolves along its unique phase space trajectory. We refer to this formal procedure as a \textit{contact reduction by scaling symmetry} \cite{bravetti2023scaling}, and note that the quotient space is no longer symplectic, but inherits a contact structure. Such contact theories are precisely of the action-dependent nature discussed above. The physical interpretation of this action dependence is that the resulting theory is non-conservative in nature.
\section{Symmetry Reduction of Multisymplectic Field Theories}\label{Sec:ReductionofMultisymplecticFieldTheory}
In order to establish notational conventions, we begin with an account of the symmetry reduction of Lagrangian systems that admit dynamical similarities. A more detailed exposition of these ideas, together with their extension to the Hamiltonian formalism, may be found in \cite{bell2025dynamical}. Consider the fibre bundle $\pi:E\rightarrow M$, over the $d$-dimensional, orientable spacetime manifold $M$, with local coordinates $x^{\mu}$, and volume form $\text{vol}=dx^0\wedge\,\cdots\,\wedge dx^{d-1}:=d^dx$. The covariant configuration space $E$ is of dimension $(d+n)$, and the first jet bundle $\kappa:J^1E\rightarrow E$ of sections of $\pi$ corresponds to the velocity phase space. We introduce the bundle projection $\widehat{\pi}:= \pi\circ \kappa:J^1E\rightarrow M$, taking local coordinates on $J^1E$ to be $(x^{\mu},y^a,y^a_{\mu})$, with $1\leqslant a \leqslant n$. From this, it is clear that $J^1E$ is of dimension $d+n(1+d)$. The Lagrangian density $\mathcal{L}$ is a $d$-form on $J^1E$, satisfying $\iota_V\mathcal{L}=0$ for any $\widehat{\pi}$-vertical vector field $V$. In local coordinates, we write
\begin{equation}\label{Eq:LagrangianDensity}
    \mathcal{L}(x^{\mu},y^a,y^a_{\mu}) = L(x^{\mu},y^a,y^a_{\mu})\,\widehat{\pi}^*\text{vol}
\end{equation}  
in which $L:J^1E \rightarrow \mathbb{R}$ denotes the Lagrangian function. The (pre-)multisymplectic form $\Omega_L\in \Omega^{d+1}(J^1E)$ is defined to be $\Omega_L:=-d\Theta_L$, with
\begin{equation}\label{Eq:CartanForms}
    \Theta_L = \frac{\partial L}{\partial y^a_{\mu}}\,dy^a\wedge d^{d-1}x_{\mu} - \left(\frac{\partial L}{\partial y^a_{\mu}}\,y^a_{\mu} - L\right)\,d^dx
\end{equation}
For the purpose of implementing a contact reduction by scaling symmetry, it will be of benefit to introduce
\begin{equation}\label{Eq:Lagrangian1Forms}
    \theta^{\mu}_L := -\, \iota_{\partial_{d-1}}\,\cdots \,\,\iota_{\partial_0} (\Theta_{L} \wedge dx^{\mu}) = \frac{\partial L}{\partial y^{\mu}_a}\,dy^a
\end{equation}
Given a multisymplectic Lagrangian system $(J^1E,\Omega_L)$, the variational problem determines critical sections $\phi\in\Gamma(M,E)$, whose canonical lifting $j^1\phi$ to $J^1E$ satisfy the Euler-Lagrange field equations. In local coordinates, if $\phi(x)=\left(x^{\mu},y^a(x)\right)$, then $j^1\phi(x)=\left(x^{\mu},y^a(x),\frac{\partial y^a}{\partial x^{\mu}}(x)\right)$, and we have
\begin{equation}\label{Eq:MultisymplecticEOM2}
    \frac{\partial}{\partial x^{\mu}}\,\left(\frac{\partial L}{\partial y^a_{\mu}}\circ j^1\phi\right) - \frac{\partial L}{\partial y^a}\circ j^1\phi = 0
\end{equation}
From the forms (\ref{Eq:Lagrangian1Forms}), we identify a vector field $\Sigma\in\mathfrak{X}^{\infty}(J^1E)$ as a scaling symmetry of degree $\Lambda\in\mathbb{R}$ if
\begin{equation}\label{Eq:ScalingSymmetry}
    \mathfrak{L}_{\Sigma}L=\Lambda L\hspace{3cm}\mathfrak{L}_{\Sigma}\theta^{\mu}_L=\theta^{\mu}_L \quad\quad\textrm{for}\;\mu=0,\,\cdots,d-1
\end{equation}
By means of a suitable change of variables, we adopt coordinates on $J^1E$ adapted to the scaling direction, so that
\begin{equation}
    \Sigma=\xi\frac{\partial}{\partial \xi} + \xi_{\mu}\frac{\partial}{\partial \xi_{\mu}}
\end{equation}
In what follows, we shall focus on those cases in which $\xi$ is a covariant configuration space variable: that is, a coordinate on $E$. Such a requirement is not overly restrictive, and allows us to be more general about the geometrical interpretation of the reduced system. Moreover, this is the case we shall encounter with the Palatini action. Defining $e^{\rho/\Lambda}=\xi$, we find that
\begin{equation}\label{Eq:LandSigma}
    L=e^{\rho}f(\rho_{\mu},\phi^a,\phi^a_{\mu}) \hspace{3cm} \Sigma = \Lambda\frac{\partial}{\partial \rho}
\end{equation}
in which $\phi^a$ are fields left unscaled by $\Sigma$, and $\phi^a_{\mu}$ their corresponding velocities \cite{bell2025dynamical}. We now introduce
\begin{equation}\label{Eq:sandL^H}
    s^{\mu}:=\frac{\partial f}{\partial \rho_{\mu}} \hspace{3cm} L^H:= f-\rho_{\mu}s^{\mu}
\end{equation}
The coordinate $\rho$ satisfies $\rho_{\mu}=-\frac{\partial L^H}{\partial s^{\mu}}$, and the variables $s^{\mu}$ are components of the action density, and confer frictional properties to the field theory described by $L^H$ \cite{sloan2021scale}.\\

In general, multicontact (or Herglotz) Lagrangians have a similar geometrical description to that of the multisymplectic objects discussed so far. From the velocity phase space $J^1E$, we define
\begin{equation}\label{Eq:MulticontactConfigurationBundle}
    \mathcal{S} := J^1E \times_M \wedge^{d-1}T^*M \cong J^1E \times \mathbb{R}^d
\end{equation}
with projections $\tau:\mathcal{S}\rightarrow E$ and $\beta=\pi\circ\tau:\mathcal{S}\rightarrow M$. Local coordinates on $\mathcal{S}$ are $(x^{\mu},y^a,y^a_{\mu},s^{\mu})$, and a multicontact Lagrangian density is expressed as
\begin{equation}\label{Eq:ContactLagrangian}
    \mathcal{L}(x^{\mu},y^a,y^a_{\mu},s^{\mu}) = L(x^{\mu},y^a,y^a_{\mu},s^{\mu})\,\beta^*\text{vol}
\end{equation}
In analogy to (\ref{Eq:CartanForms}), there exist forms $\Theta_L\in\Omega^d(\mathcal{S})$ and $\Omega_L\in\Omega^{d+1}(\mathcal{S})$, with
\begin{equation}\label{Eq:LagrangianForm}
    \Theta_L = \left( ds^{\mu} - \frac{\partial L}{\partial y^a_{\mu}}dy^a\right) \,\wedge \,d^{d-1}x_{\mu} + \left(\frac{\partial L}{\partial y^a_{\mu}}\,y^a_{\mu} - L\right)\,d^dx
\end{equation}
\vspace{-0.3cm}
\begin{equation}\label{Eq:ContactOmega}
    \Omega_L = d\Theta_L - \frac{\partial L}{\partial s^{\mu}}\,dx^{\mu} \wedge\,\Theta_L
\end{equation}
On $\mathcal{S}$, holonomic sections $\Psi:M\rightarrow \mathcal{S}$ is may be expressed locally as 
\begin{equation*}
    \Psi(x) = \left(x^{\mu},y^a(x), \frac{\partial y^a}{\partial x^{\mu}}\biggr|_x,s^{\mu}(x)\right)
\end{equation*}
and the multicontact analogue of the Euler-Lagrange field equations (\ref{Eq:MultisymplecticEOM2}) become
\begin{equation}\label{Eq:HerglotzLagrangeEquations}
    \begin{split}
        \frac{\partial}{\partial x^{\mu}}\left(\frac{\partial L}{\partial y^a_{\mu}}\circ \Psi\right) = \left( \frac{\partial L}{\partial y^a}+\frac{\partial L}{\partial y^a_{\mu}}\frac{\partial L}{\partial s^{\mu}}\right) \circ \Psi\\
        \frac{\partial s^{\mu}}{\partial x^{\mu}} = L\circ \Psi
    \end{split}
\end{equation}
These ideas must be modified slightly when the multicontact Lagrangian is obtained via the elimination of a scaling variable. In particular, the change of variables $e^{\rho/\Lambda}=\xi$ required to render the multisymplectic Lagrangian of the form (\ref{Eq:LandSigma}) clearly does not cover all of $J^1E$, since $e^{\rho/\Lambda}>0$. Mathematically, $\xi$ has been identified as a globally-defined object satisfying $\mathfrak{L}_{\Sigma}\xi=\xi$; such quantities are referred to as scaling functions. By writing $\xi=e^{\rho/\Lambda}$, we have revealed that the configuration space separates into two connected pieces $E_{\pm}$, with $E_{\pm}\cong \tilde{E}\times_M\left(M\times\mathbb{R}_{\pm}\right)$. Here, $\tilde{E}$ is a codimension-1 subspace of $E$, interpreted as a configuration space composed of all field variables of $E$, except $\xi$. The trivial bundles $M\times \mathbb{R}_{\pm}\rightarrow M$ correspond to the scaling variable, taking either $\xi=+\,e^{\rho/\Lambda}$ or $\xi=-\,e^{\rho/\Lambda}$. Provided both components are accounted for, we may study a dynamically-equivalent theory, defined on the space $\mathcal{S}:=J^1\tilde{E}\times \mathbb{R}^d$.\\

The Herglotz Lagrangian (\ref{Eq:sandL^H}) is a function on $J^1\tilde{E}\times \mathbb{R}^d$ and \textit{not} on $J^1\tilde{E}$ precisely because the scaling variable has been eliminated, whilst the $d$ velocity coordinates $\rho_{\mu}$ have not. When $(J^1E,\Theta_L)$ is a regular system, these coordinates - which, in the reduced theory, assume the role of the action density - generate a Reeb distribution, with the result that\footnote{Here, $\Theta_{L^H}$ refers to the $d$-form on the reduced space, calculated according to (\ref{Eq:LagrangianForm}).} $(\mathcal{S},\Theta_{L^H})$ is a regular multicontact Lagrangian system, whose dynamics are deduced from (\ref{Eq:HerglotzLagrangeEquations}).
\section{Orthonormal Frames and Spin Connections}\label{Sec:OrthonormalFrames}
Given a $d$-dimensional manifold $M$, the frame bundle $Fr(TM):=FM\rightarrow M$ is a principal $GL(\mathbb{R}^d)$-bundle, whose elements are pairs $(x,\Delta_x)$, where $\Delta_x$ denotes a frame at the point $x\in M$. This frame is a set of $d$ linearly independent sections $\Delta_x = (e_1|_x,\,\cdots,e_d|_x)$, which provide a basis for the tangent space $T_xM$ \cite{nakahara2018geometry}. At each point, the tangent space admits two distinct bases: that provided by the elements $e_I$ of the frame, and that of the coordinate derivatives $\partial_{\mu}$. Throughout, $\mu,\nu$ are spacetime indices, while $I,J$ denote internal indices. There exist components $e_{\mu}^{\;I}(x)$ of a $GL(\mathbb{R}^d)$ matrix, such that $\partial_{\mu}=e_{\mu}^{\;I}(x)\,e_I$, and for some $v\in T_xM$, we may interchangeably write $v=v^\mu \partial_\mu$ or $v=v^Ie_I$. An analogous statement holds for the cotangent bundle $T^*M$: at each point $x\in M$, we have a coordinate basis of 1-forms $dx^{\mu}$, and a basis provided by the elements $e^I$ of the coframe.\\

Consider the case of $d=4$, and introduce a Lorentzian metric $g_{\mu\nu}$ of signature $(-,+,+,+)$ on $M$. When $M$ is orientable, there is a reduction of the structure group of $FM$ from $GL(\mathbb{R}^4)$ to $SO(1,3)$, upon identifying the orientable orthonormal frames as a preferred subset. These satisfy $g(e_I,e_J) = \eta_{IJ}$, and we denote the subbundle of orientable orthonormal frames $SO(M)\subset FM$ \cite{baez1994gauge}, constructing the associated bundle
\begin{equation}\label{Eq:AssociatedFrameBundle}
    E = SO(M) \times_{SO(1,3)}\mathbb{R}^{1,3}
\end{equation}
in which we take the fundamental representation of $SO(1,3)$. The orthornormal frame bundle $SO(M)\rightarrow M$ is a principal bundle, which admits a connection; this may be represented by a (unique) Lie algebra valued 1-form $\omega\in\Omega^1(SO(M),\mathfrak{so}(1,3))$. Given some trivialising section $s:U\subset M \rightarrow SO(M)$, the quantity $s^*\omega\in \Omega^1(U,\mathfrak{so}(1,3))$ is a local spin connection 1-form, which we denote $\omega^{IJ}$. This connection naturally induces a corresponding structure on the associated bundle $E$, giving a covariant derivative $D$, whose action on some $X=X^Ie_I$ reads
\begin{equation}\label{Eq:SpinCovariantDerivative}
    DX^I = dX^I + \omega^I_{\;J}\wedge X^J
\end{equation}
While $D$ acts on internal indices, there is also a connection $\nabla$ on the tangent bundle, which is sensitive only to spacetime indices
\begin{equation}\label{Eq:NablaConnection}
    \nabla_{\mu}V^{\nu} = \partial_{\mu} V^{\nu} +\Gamma^{\nu}_{\mu\lambda} V^{\lambda}
\end{equation}
A priori, $D$ and $\nabla$ are completely independent, and neither is required to be metric-compatible. However, one obtains more interesting outcomes when additional conditions are imposed. Each connection has its own notion of torsion:
\begin{equation}\label{Eq:Torsion1}
    T_{\mu\nu}^{\;\;\;\lambda}:=\Gamma_{\mu\nu}^{\lambda}-\Gamma_{\nu\mu}^{\lambda} \hspace{2cm} T^I:= de^I+ \omega^I_{\;K}\wedge\,e^K
\end{equation}
Finally, we introduce a `total covariant derivative' $\mathcal{D}$, which acts on both spacetime and internal indices:
\begin{equation}\label{Eq:DExtension}
    \mathcal{D}_{\mu} V^{\lambda I} := \partial_{\mu}X^{\lambda I} + \Gamma^{\lambda}_{\mu\rho}V^{\rho I} + \omega_{\mu\;\;J}^{\;\;I} \, V^{\lambda J}
\end{equation}
\section{Gravitational Actions}\label{Sec:GravActions}
The conventional way in which one formulates a gravitational action is to employ the unique, torsion-free, metric-compatible connection, acting on sections of the tangent bundle. The resulting Lagrangian is a function of the metric tensor, together with its first and second derivatives \cite{wald2010general}. This second-order formalism is not particularly amenable to a treatment based on multisymplectic geometry, as we would be required to work on the \textit{second} jet bundle. For this reason, we shall favour the first-order formalism, and throughout, we follow the conventions of \cite{peldan1994actions} with respect to index (anti)symmetrisation, taking $A^{(\mu\nu)} := A^{\mu\nu} + A^{\nu\mu}$ and $A^{[\mu\nu]} := A^{\mu\nu} - A^{\nu\mu}$.\\

Within the second-order framework, the tangent bundle connection is assumed metric-compatible ($\nabla_{\alpha}g_{\mu\nu}=0$) and torsion-free ($T^{\;\;\;\lambda}_{\mu\nu}=0$). The total covariant derivative $\mathcal{D}$ is also compatible with the spacetime metric ($\mathcal{D}_{\alpha}g_{\mu\nu}=0)$, and the internal Minkowski metric ($\mathcal{D}_{\mu}\eta_{IJ}=0$). This latter condition implies that $\omega_{\mu IJ}$ (and by extension $\omega_{\mu}^{\;\;IJ}$) is antisymmetric in its Lie algebra indices. The above conditions are common to both the first and second-order formalisms. The distinguishing feature of a second-order action is the tetrad postulate
\begin{equation}\label{Eq:CompatibilityCondition}
    \mathcal{D}_{\mu}e_{\nu}^{\;I} = \partial_{\mu}e_{\nu}^{\;I} - \Gamma^{\lambda}_{\mu\nu} e_{\lambda}^{\;I} + \omega_{\mu\;\;J}^{\;\;I}\,e_{\nu}^{\;J} \stackrel{!}{=}0
\end{equation}
This condition identifies a unique, torsion-free spin connection, compatible with the frame field, which may be expressed as
\begin{equation}\label{Eq:OmegaTetrads}
    \omega_{\mu}^{\;\;IJ} = \frac{1}{2}e^{\lambda[I}\left(\partial_{[\mu}e_{\lambda]}^{\;\;J]}+ e^{\nu J]}e_{\mu}^{\;K}\partial_{\nu} e_{\lambda K}\right)
\end{equation}
The curvature of the tangent bundle connection is an $\textrm{End}(TM)$ valued 2-form, which we denote $r$, so as to reserve $R$ for that of the $SO(1,3)$ connection
\begin{equation}\label{Eq:NablaCurvature}
    \nabla_{[\mu}\nabla_{\nu]} V_{\lambda} = r_{\mu\nu\lambda}^{\quad\;\;\rho} \; V_{\rho} + T_{\mu\nu}^{\;\;\;\rho}\,\nabla_{\rho}V_{\lambda}
\end{equation}
For completeness, we have provided the generalisation for non-zero torsion. The curvature of the spin connection is an $\mathfrak{so}(1,3)$ valued 2-form. Since $\mathcal{D}$ may act on both internal and spacetime indices, we have
\begin{equation}\label{Eq:Curvatures}
    \mathcal{D}_{[\mu}\mathcal{D}_{\nu]}V_{\lambda} = \, R_{\mu\nu\lambda}^{\quad\;\;\rho} \; V_{\rho}+ T_{\mu\nu}^{\;\;\;\rho}\, \mathcal{D}_{\rho}V_{\lambda}\quad\quad\quad\quad \mathcal{D}_{[\mu}\mathcal{D}_{\nu]}V_{I} = R_{\mu\nu I}^{\quad\;\;J} \; V_J + T_{\mu\nu}^{\;\;\;\rho}\, \mathcal{D}_{\rho}V_{I}
\end{equation}
From a direct calculation with $T_{\mu\nu}^{\;\;\;\rho}=0$, we see that
\begin{equation}\label{Eq:CoordinateExpressionCurvature}
    R_{\mu\nu}^{\quad IJ}= \partial_{[\mu}\omega_{\nu]}^{\;\;\; IJ} + \omega_{[\mu}^{\;\;\; IK}\omega_{\nu]K}^{\quad\;\;J}
\end{equation}
and the Ricci scalar $R$ reads
\begin{equation}\label{Eq:FRicciScalar}
    R=g^{\mu\nu}R_{\mu\nu} = e^{\mu}_{\;I} e^{\nu}_{\;J} \,R_{\mu\nu}^{\quad IJ}
\end{equation}
As a consequence of the tetrad postulate, the second-order Einstein-Hilbert action takes the frame field as its sole dynamical variable. Relaxing the assumption that $\mathcal{D}_{\mu}e_{\nu}^{\;I}=0$, we obtain a theory of two independent variables, in which the spin connection is no longer required to be torsion-free. This is precisely the content of the first-order formalism, for which the corresponding Hilbert-Palatini action is written as
\begin{equation}\label{Eq:HPAction1}
    S_{\textrm{HP}}\left[e,\omega\right] = \frac{1}{2\kappa}\int d^4x\;e\,e^{\mu}_{\;I}e^{\nu}_{\;J} R_{\mu\nu}^{\quad IJ}
\end{equation}
The variation of this action is a standard calculation \cite{ashtekar1991lectures,ciaglia2023geometry}, and so we shall not belabour the point excessively; however, variation with respect to $\omega_{\mu}^{\;\;IJ}$ is illustrative, as it requires partial integration, which is relatively non-trivial for curved manifolds with torsion \cite{yoon2019lagrangian,penrose1984spinors}. Making use of
\begin{equation}
    \delta e = -\, e\,e_{\mu}^{\;I}\delta e^{\mu}_{\;I}
\end{equation}
variations with respect to the frame field read 
\begin{equation*}
      \delta S_{\textrm{HP}} = \frac{1}{2\kappa}\int d^4x \; e\,\biggr(2 e^{\mu}_{\;K} e^{\lambda}_{\;I}e^{\nu}_{\;J}R_{\lambda\nu}^{\quad KJ} - e^{\mu}_{\;I}e^{\lambda}_{\;K} e^{\rho}_{\;L}\,R_{\lambda\rho}^{\quad KL} \biggr) \,\delta e_{\mu}^{\;I}  
\end{equation*}
and since $g_{\mu\nu}=e_{\mu}^{\;I}e_{\nu}^{\;J}\eta_{IJ}$, the bracketed term is easily manipulated into the familiar form $R_{\mu\nu}-\frac{1}{2}Rg_{\mu\nu}=0$. While identical in form to the vacuum field equations, $e^I$ and $\omega^{IJ}$ are \textit{independent}, and so the curvature $R$ is that of the gauge connection, and \textit{not} the tangent bundle. The second component of the calculation requires that we independently vary the spin connection. To this end, note that, with $\mathcal{D}_{\alpha}g_{\mu\nu}=0$, we may use
\begin{equation}\label{Eq:IBP}
    \partial_{\mu}\left(eA^{\mu}B^{\nu}C_{\nu}\right) = e\left(\mathcal{D}_{\mu}A^{\mu}\right)B^{\nu}C_{\nu} + eA^{\mu}\mathcal{D}_{\mu}\left(B^{\nu}C_{\nu}\right) + e\,T_{\rho\mu}^{\;\;\;\rho}\,A^{\mu}B^{\nu}C_{\nu}
\end{equation}
Setting the torsion to zero, and noting that variations of $\omega_{\mu}^{\;\;IJ}$ give
\begin{eqnarray}
    \delta R_{\mu\nu}^{\quad IJ}= \partial_{[\mu}\delta\omega_{\nu]}^{\;\;\; IJ} + \delta \omega_{[\mu}^{\;\;IK}\omega_{\nu]K}^{\quad\;\;J} + \omega_{[\mu}^{\;\;\;IK}\delta\omega_{\nu]K}^{\quad\;\;J} = \mathcal{D}_{[\mu}\delta\omega_{\nu]}^{\quad IJ}
\end{eqnarray}
we have
\begin{equation*}
    \delta S_{\textrm{HP}} = \frac{1}{2\kappa}\int d^4x\;e\,e^{\mu}_{\;I}e^{\nu}_{\;J}\,\mathcal{D}_{[\mu}\delta\omega_{\nu]}^{\quad IJ} = -\, \frac{1}{2\kappa}\int d^4x\;e\,\mathcal{D}_{\mu}(e^{[\mu}_{\;I}e^{\nu]}_{\;J})\delta\omega_{\nu}^{\;\;\;IJ}+ \frac{1}{2\kappa}\int d^4x \,\partial_{\mu}\left(e\,e^{[\mu}_{\;I}e^{\nu]}_{\;J}\,\delta\omega_{\nu}^{\;\;\; IJ}\right)
\end{equation*}
Discarding the total divergence, we have
\begin{equation}\label{Eq:HPEoM}
    R_{\mu\nu}-\frac{1}{2}Rg_{\mu\nu}=0,\quad\quad\quad \mathcal{D}_{\mu}(e^{[\mu}_{\;I}e^{\nu]}_{\;J})=0
\end{equation}
Compatibility of the frame field is thus enforced dynamically, and implies that $R_{\mu\nu}$ and $R$ coincide with the corresponding objects derived from the tangent bundle connection.\\

Thus far, we have presented the standard geometrical framework of the first-order action. In order to make contact with the ideas of section (\ref{Sec:ReductionofMultisymplecticFieldTheory}), we introduce the following covariant configuration space \cite{gaset2018new,fatibene2013natural}
\begin{equation}\label{Eq:CovarConf}
    \mathcal{E} := \left(T^*M \otimes E\right) \times_M C(SO(M))
\end{equation}
in which $C(SO(M))\rightarrow M$ is the affine bundle of connections over $M$, modelled on the vector bundle $T^*M\times\textrm{Ad}(SO(M))\rightarrow M$. Local coordinates on $J^1\mathcal{E}$ are then $(x^{\mu},e_{\mu}^{\;I},\omega_{\mu}^{\;\;IJ},\partial_{\mu} e_{\nu}^{\;I}\,,\partial_{\mu}\omega_{\nu}^{\;\;IJ})$, and we introduce the following projections
\begin{equation*}
    \Pi:E\rightarrow M \quad\quad\quad \pi:\mathcal{E}\rightarrow M \quad\quad\quad \kappa: J^1\mathcal{E}\rightarrow \mathcal{E} \quad\quad\quad \widehat{\pi} = \pi\circ \kappa :J^1\mathcal{E}\rightarrow M
\end{equation*}
The Lagrangian density is a 4-form on $J^1\mathcal{E}$, given by
\begin{equation}\label{Eq:HPLagrangian}
    \mathcal{L}_{\textrm{HP}} = \frac{1}{2\kappa}\, e\,e^{\mu}_{\;I}e^{\nu}_{\;J}R_{\mu\nu}^{\quad IJ} \;\widehat{\pi}^*\text{vol}
\end{equation}
\section{A First-Order Frictional Action}\label{Sec:FirstOrderFrictionalAction}
In 3+1 spacetime dimensions, the metric $g$ possess ten independent components; nine of these describe the conformal geometry of the spacetime, while the tenth provides a notion of volumetric scale. It is possible to decompose $g$ in a manner which reflects this distribution of the degrees of freedom, writing $g_{\mu\nu}=e^{2\phi}G_{\mu\nu}$. The conformal factor $\phi(x)$ contains all local scale information, while $G_{\mu\nu}$ is a symmetric tensor of fixed determinant $-1$ \cite{wang2006conformaldecompositioncanonicalgeneral}. Throughout, we assume that the spacetime manifold $M$ is (time-)orientable, and admits a globally-defined, nowhere-vanishing frame field.
\subsection{Constructing the Action-Dependent Theory}\label{Subsec:Construction}
Working with the coframe $e^I$, we write $e^I=e^\phi\tilde{e}^I$, with $\textrm{det}\,\tilde{e}^I=1$, so that $e_{\mu}^{\;I} = e^{\phi} \tilde{e}_{\mu}^{\;I}$ and $e^{\mu}_{\;I} = e^{-\phi} \tilde{e}^{\mu}_{\;I}$. Consider the effect of this decomposition on the torsion $T^I$
\begin{equation}\label{Eq:Torsion2}
    T^I=d(e^{\phi}\tilde{e}^I) + \omega^I_{\;K}\wedge (e^{\phi}\tilde{e}^K) = e^{\phi}\biggr[d\tilde{e}^I + \omega^I_{\;K}\wedge \tilde{e}^K + d\phi\wedge \tilde{e}^I\biggr]
\end{equation}
On-shell, the torsion of $\omega^I_{\;J}$ is vanishing; comparison of this expression with (\ref{Eq:Torsion1}) suggests that variations of the conformally-decomposed action will \textit{not} identify $\omega^I_{\;J}$ as the connection compatible with $\tilde{e}^I$. It would be natural, therefore, to seek a similarly scale-invariant object $\tilde{\omega}^I_{\;J}$, such that, in the same way that $\omega^I_{\;J}$ is dynamically fixed to be compatible with $e^I$, the modified connection $\tilde{\omega}^I_{\;J}$ should have an identical property, with respect to $\tilde{e}^I$. From above, we see that $\tilde{\omega}^I_{\;J}$ should satisfy
\begin{equation}\label{Eq:NewCompatibility}
    d\tilde{e}^I + \omega^I_{\;K}\wedge \,\tilde{e}^K =-\, d\phi\wedge \tilde{e}^I \quad\quad\implies\quad\quad d\tilde{e}^I + \tilde{\omega}^I_{\;K}\wedge\, \tilde{e}^K = 0
\end{equation}
A short calculation in local coordinates confirms that
\begin{equation}\label{Eq:TransformedConnection}
    \tilde{\omega}_{\mu}^{\;\;IJ} = \omega_{\mu}^{\;\;IJ} + \tilde{e}^{\lambda [I} \tilde{e}_{\mu}^{\;J]}\,\partial_{\lambda}\phi
\end{equation}
satisfies this requirement, and crucially, retains antisymmetry in its upper indices $I$ and $J$. Expressing the action (\ref{Eq:HPAction1}) in terms of the variables $(\phi,\tilde{e}^I,\tilde{\omega}^I_{\;J})$, we find
\begin{equation}\label{Eq:ConformalAction1}
    S = \frac{1}{2\kappa}\int d^4x\;e^{2\phi}\biggr[\tilde{e}^{\mu}_{\;I}\tilde{e}^{\nu}_{\;J} \left( \tilde{R}_{\mu\nu}^{\quad IJ} - \tilde{\mathcal{D}}_{[\mu}\left( \tilde{e}^{\lambda[I} \tilde{e}_{\nu]}^{\;\;J]} \partial_{\lambda}\phi\right) \right)- 6 \tilde{e}^{\mu K} \tilde{e}^{\nu}_{\;K} \partial_{\mu}\phi \, \partial_{\nu} \phi \biggr]
\end{equation}
in which we have defined 
\begin{equation*}
    \tilde{R}_{\mu\nu}^{\quad IJ}:= \partial_{[\mu}\tilde{\omega}_{\nu]}^{\;\;\; IJ} + \tilde{\omega}_{[\mu}^{\;\;\; IK}\tilde{\omega}_{\nu]K}^{\quad\;\;J}
\end{equation*}
and $\tilde{\mathcal{D}}$ is a covariant derivative which acts only on $SO(1,3)$ indices, utilising the conformal spin connection; explicitly, for a tensor of mixed indices $A_{\lambda}^{IJ}$
\begin{equation*}
    \tilde{\mathcal{D}}_{\mu}A_{\lambda}^{IJ}:= \partial_{\mu}A_{\lambda}^{IJ} + \tilde{\omega}_{\mu\;\;K}^{\;\;I}A_{\lambda}^{KJ} + \tilde{\omega}_{\mu\;\;K}^{\;\;J}A_{\lambda}^{IK}
\end{equation*}
At present, this action is not amenable to a contact reduction, due to the $\partial_{\lambda}\phi$ inside the covariant derivative; however, integrating by parts, discarding surface terms, and writing the conformal factor as $\Phi:=2\phi$, we find 
\begin{equation}\label{Eq:ConformalAction2}
    S = \frac{1}{2\kappa}\int d^4x \;e^{\Phi}\biggr[\tilde{e}^{\mu}_{\;I}\tilde{e}^{\nu}_{\;J}\tilde{R}_{\mu\nu}^{\quad IJ} + \tilde{e}^{\lambda[I} \tilde{e}_{\nu}^{\;\;J]} \partial_{\lambda}\Phi\, \tilde{\mathcal{D}}_{\mu}\left(\tilde{e}^{\mu}_{\;I}\tilde{e}^{\nu}_{\;J}\right) + \frac{3}{2} \tilde{e}^{\mu K} \tilde{e}^{\nu}_{\;K} \partial_{\mu}\Phi \, \partial_{\nu} \Phi \biggr]
\end{equation}
From the results of section (\ref{Sec:ReductionofMultisymplecticFieldTheory}), the action is now of a form to which a contact reduction may be applied. Indeed, calculation of the forms $\theta^\mu_L$ reveals that $\mathfrak{L}_{\partial_\Phi}\theta_L^\mu=\theta_L^\mu$, and so $\Sigma=\partial_{\Phi}$ is a scaling symmetry of degree one. The Lagrangian 
\begin{equation}\label{Eq:LagrangianFunction}
    L(x^{\mu},\tilde{e}^I,\tilde{\omega}^I_{\;J},\Phi) = \frac{1}{2\kappa} e^{\Phi} \left( \tilde{e}^{\mu}_{\;I}\tilde{e}^{\nu}_{\;J}\tilde{R}_{\mu\nu}^{\quad IJ} + \tilde{e}^{\lambda[I} \tilde{e}_{\nu}^{\;\;J]} \partial_{\lambda}\Phi\, \tilde{\mathcal{D}}_{\mu}\left(\tilde{e}^{\mu}_{\;I}\tilde{e}^{\nu}_{\;J} \right) + \frac{3}{2} \, \tilde{e}^{\mu K} \tilde{e}^{\nu}_{\;K} \partial_{\mu}\Phi \, \partial_{\nu} \Phi \right)
\end{equation}
is manifestly of the form $L=e^{\Phi}f$, and so we have
\begin{equation}\label{Eq:ActionDensity}
    s^{\mu} = \frac{\partial f}{\partial (\partial_{\mu}\Phi)} =\frac{1}{2\kappa} \biggr[ 3\tilde{e}^{\mu K} \tilde{e}^{\nu}_{\;K} \partial_{\nu}\Phi + \tilde{e}^{\mu[I}\tilde{e}_{\nu}^{\;J]}\tilde{\mathcal{D}}_{\rho}\left(\tilde{e}^{\rho}_{\;I}\tilde{e}^{\nu}_{\;J}\right)\biggr]
\end{equation}
The Herglotz Lagrangian is constructed as $L^H = f-(\partial_{\mu}\Phi ) s^{\mu}$, and by inverting the above expression, we eliminate $\partial_{\mu}\Phi$ in favour of the action density. This calculation is somewhat lengthy, and offers little physical insight; we therefore omit the details, and affirm that the Herglotz Lagrangian has the following form
\begin{equation}\label{Eq:HerglotzLagrangian}
    \begin{split}
        L^H = \frac{1}{2\kappa} \tilde{e}^{\mu}_{\;I}\tilde{e}^{\nu}_{\;J}\tilde{R}_{\mu\nu}^{\quad IJ} - \frac{\kappa}{3} \, \tilde{e}_{\mu}^{\;K} \tilde{e}_{\nu K} s^{\mu}s^{\nu} + \frac{1}{3} s^{\mu} \tilde{e}_{\mu}^{\;[I}\tilde{e}_{\lambda}^{\;J]}\tilde{\mathcal{D}}_{\sigma}(\tilde{e}^{\sigma}_{\;I}\tilde{e}^{\lambda}_{\;J})\\
        - \frac{1}{12\kappa}\,\tilde{e}_{\mu}^{\;[I}\tilde{e}_{\rho}^{\;J]}\,\tilde{e}^{\mu[A}\tilde{e}_{\lambda}^{\;B]}\,\tilde{\mathcal{D}}_{\sigma}\left(\tilde{e}^{\sigma}_{\;I}\tilde{e}^{\rho}_{\;J}\right)\tilde{\mathcal{D}}_{\alpha}\left(\tilde{e}^{\alpha}_{\;A}\tilde{e}^{\lambda}_{\;B}\right)
    \end{split}
\end{equation}
The first term of this Lagrangian is clearly a scale-free version of the original Palatini theory, while the second is quadratic in the action density. While the conformal factor \textit{is} an empirically inaccessible degree of freedom, it does still possess an algebraic status within the Palatini action. Its excision does not, therefore, leave the rest of the Lagrangian unaffected. Indeed, as is ubiquitously the case with contact reduction, we eliminate a scaling degree of freedom at the cost of introducing friction, in the form of action-dependence. The remaining terms in (\ref{Eq:HerglotzLagrangian}) contain covariant derivatives of the frame fields, which vanish on-shell, as is shown in appendix \ref{App:Appendix1}. From the results of section (\ref{Sec:ReductionofMultisymplecticFieldTheory}), we know that the Herglotz Lagrangian (\ref{Eq:HerglotzLagrangian}), together with the condition $\partial_{\mu}s^{\mu}=L^H$, faithfully reproduces the complete dynamical content of the original Palatini action, without reference to the scaling variable $\phi$. We postpone further discussion of the equations of motion until section (\ref{Sec:Metric}).\\

While the Herglotz Lagrangian (\ref{Eq:HerglotzLagrangian}) is relatively unwieldy, the corresponding density may be cast exclusively in terms of forms, and is somewhat less complex
\begin{equation}\label{Eq:LHForms}
    \mathcal{L}^H = \frac{1}{2\kappa}\star\Sigma^{IJ}\wedge \, \tilde{R}_{IJ} - \frac{\kappa}{3}\left( s\wedge\star\, s\right) + \frac{1}{3}\,(\star\,s)\,\wedge\, \star\, \iota_{\Sigma_{IJ}}\tilde{\mathcal{D}}\Sigma^{IJ}  - \frac{1}{12\kappa} \,\iota_{\Sigma_{IJ}}\tilde{\mathcal{D}}\Sigma^{IJ} \wedge\star\; \iota_{\Sigma_{AB}}\tilde{\mathcal{D}}\Sigma^{AB}
\end{equation}
Here, we have defined $\Sigma_{IJ}:=\tilde{e}_I\wedge\, \tilde{e}_J$, and $\tilde{R}_{IJ}:=\frac{1}{2}\tilde{R}_{\mu\nu IJ} \, dx^{\mu}\wedge\, dx^{\nu}$. The Hodge dual $\star$ should be understood to be taken with respect to the $\tilde{e}^I$.\\

Having deduced this form of $\mathcal{L}^H$, it is possible to relax the assumption of the independence of $\tilde{e}^I$ and $\tilde{\omega}^I_{\;J}$. The result is a second-order frictional Lagrangian, which may be used as a starting point for performing calculations in General Relativity, without reference to the conformal factor
\begin{equation}\label{Eq:MainResult}
    \boxed{\mathcal{L}^H = \frac{1}{2\kappa}\star\Sigma^{IJ}\wedge \, \tilde{R}_{IJ} - \frac{\kappa}{3}\left( s\wedge\star\, s\right)}
\end{equation}
Such a theoretical framework is potentially highly attractive from the perspective of quantisation schemes. Of course, our claim is by no means as bold as that of affirming that we have made progress in such a quantisation attempt. We merely observe that our reduced theory, in making no reference to $\phi$, naturally sidesteps a number of issues that arise within conventional approaches, such as the path integral.
\subsection{Geometry of the Reduced Space}\label{Subsec:GeometryofReducedSpace}
Decomposing the coframe as $e^I=e^{\phi}\tilde{e}^I$, with $\text{det}\,\tilde{e}^{I}=1$, implies that the covariant configuration bundle $\mathcal{E}$ admits the following decomposition
\begin{equation}\label{Eq:AdaptedConfigurationBundle}
    \mathcal{E} \cong (M\times \mathbb{R}_+)\times_M (\widetilde{E}\otimes T^*M)\times_M C(SO(M))
\end{equation}
Here, $\widetilde{E}$ is a codimension-1 subspace of $E$, such that sections of $\widetilde{E} \otimes T^*M$ are the coframe fields $\tilde{e}^I$, and $M\times\mathbb{R}_+\rightarrow M$ is a trivial bundle over $M$, corresponding to the scalar field $\phi$. Expressed in this way, it is clear that the quotient space under the orbit of the scaling symmetry $\Sigma=\partial_{\Phi}$ is
\begin{equation*}
    \mathcal{E}_{\textrm{red}} \cong  (\widetilde{E}\otimes T^*M)\times_M C(SO(M))
\end{equation*}
As discussed extensively in \cite{bell2025dynamical} and \cite{Bell:2025ooq}, the elimination of the scaling variable does \textit{not} remove the $d$ independent velocity fields $\partial_{\mu}\Phi$, which assume the role of the action density $s^{\mu}$. The Herglotz Lagrangian is thus a function on the space $J^1\mathcal{E}_{\textrm{red}} \times \mathbb{R}^4$, with local coordinates
\begin{equation*}
    (x^{\mu},\tilde{e}_{\mu}^{\;I},\tilde{\omega}_{\mu}^{\;\;IJ},\partial_{\mu}\tilde{e}_{\nu}^{\;I},\partial_{\mu}\tilde{\omega}_{\nu}^{\;\;IJ},s^{\mu})    
\end{equation*}
Referring to the discussion and notation of section (\ref{Sec:ReductionofMultisymplecticFieldTheory}), when one carries out a contact reduction, the variable $\xi$ typically assumes both positive and negative values. This requires particular care to be taken when writing $\xi=e^{\rho/\Lambda}$. If we do not also consider $\xi=-\,e^{\rho/\Lambda}$, some of the dynamical information of the original system is lost. Because $M$ is equipped with a Lorentzian metric of signature $(-,+,+,+)$, $\textrm{det}\,g<0$ at all points, and so the decomposition $e^I=e^{\phi}\tilde{e}^I$ suffers no such $\pm$ ambiguities.\\

Under a change of local coordinates $x^{\mu}\rightarrow y^{\mu}(x)$, the coframe field $\tilde{e}_\mu^{\;I}$ transforms covariantly. Any local coordinate transformation \textit{must} preserve the condition $\textrm{det}\,\tilde{e}^{I}=1$, and so we see that the gauge group is reduced from $\textrm{Diff}(M)$ to the subgroup $\textrm{Diff}_{\mu}(M)$ of unimodular diffeomorphisms. This unimodularity will reappear when considering the weak-field limit, and indeed more generally when analysing the significance of the structure of our theory.
\section{Recovering the Metric Formalism}\label{Sec:Metric}
Thus far, our analysis has been conducted within the first-order formalism, expressed in a non-coordinate basis. This choice is well-motivated: the first-order construction is required to work on the first jet bundle $J^1\mathcal{E}$. The use of $(e^I,\omega^{IJ})$ offers an intuitive geometrical interpretation of the symmetry reduction, and accommodates the addition of fermionic degrees of freedom. On the other hand, for practical applications, working with a non-coordinate basis is cumbersome; expressions quickly become unwieldy, and the algebra lengthy and tedious. It will therefore be of use to recast the results obtained into a more familiar form, employing the full spacetime metric. To this end, we write $g_{\mu\nu}=e^{\Phi} G_{\mu\nu}$, with $\textrm{det}\,G_{\mu\nu}=-\,1$. On-shell, the connection compatible with $g$ is that of Levi-Civita, whose corresponding connection coefficients are the Christoffel symbols. With the conformal decomposition of the metric, these read
\begin{align*}
    \Gamma^{\rho}_{\mu\nu} = \frac{1}{2}G^{\alpha\rho}\left(\partial_{\nu}G_{\alpha\mu}+\partial_{\mu}G_{\nu\alpha}-\partial_{\alpha}G_{\mu\nu} + G_{\alpha\mu}\partial_{\nu}\Phi + G_{\nu\alpha}\partial_{\mu}\Phi - G_{\mu\nu}\partial_{\alpha}\Phi\right)
\end{align*}
If we want a connection $\tilde{\nabla}$, compatible with $G_{\mu\nu}$ in the same way $\nabla$ is compatible with $g_{\mu\nu}$, it would make sense to write
\begin{equation}
    \tilde{\Gamma}^{\rho}_{\mu\nu} :=\frac{1}{2}G^{\alpha\rho}\left(\partial_{\nu}G_{\alpha\mu}+\partial_{\mu}G_{\nu\alpha}-\partial_{\alpha}G_{\mu\nu}\right)
\end{equation}
with the result that
\begin{equation}\label{Eq:Gammas}
    \Gamma^{\rho}_{\mu\nu} = \tilde{\Gamma}^{\rho}_{\mu\nu} + \frac{1}{2}\left(\delta^{\rho}_{\mu}\partial_{\nu}\Phi + \delta^{\rho}_{\nu}\partial_{\mu}\Phi - G_{\mu\nu}\partial^{\rho}\Phi\right)
\end{equation}
That this is a sensible construction can be seen by considering the metric-compatibility condition $\nabla_{\alpha}g_{\mu\nu}=0$
\begin{equation*}
    \nabla_{\alpha}g_{\mu\nu}=\partial_{\alpha}g_{\mu\nu} - \Gamma^{\rho}_{\alpha\mu}g_{\rho\nu} - \Gamma^{\rho}_{\alpha\nu}g_{\mu\rho}=0 \quad\stackrel{(\ref{Eq:Gammas})}{\implies} \quad \tilde{\nabla}_{\alpha}G_{\mu\nu}=\partial_{\alpha}G_{\mu\nu} - \tilde{\Gamma}^{\rho}_{\alpha\mu}G_{\rho\nu} - \tilde{\Gamma}^{\rho}_{\alpha\nu}G_{\mu\rho}=0
\end{equation*}
This argument parallels the construction of the compatible spin connection $\tilde{\omega}^{IJ}$ in section (\ref{Sec:FirstOrderFrictionalAction}); indeed, by virtue of this analogy, it follows that the second-order Herglotz Lagrangian may be expressed as
\begin{equation}\label{Eq:HerglotzLag2}
    \boxed{L^H = \frac{1}{2\kappa}\tilde{R}(G) - \frac{\kappa}{3}G_{\mu\nu}s^{\mu}s^{\nu}}
\end{equation}
where $\tilde{R}(G)$ emphasises that the curvature is a function of $G_{\mu\nu}$, together with its first and second derivatives. This expression, in conjunction with (\ref{Eq:MainResult}), constitutes the core result of our work. We have constructed an action for GR which, as we proceed to show, despite making no reference to the conformal mode, correctly reproduces the complete dynamical content of the original theory.\\

Note that, for the connection $\tilde{\nabla}$, we have the following simplification
\begin{equation*}
    \tilde{\Gamma}^{\mu}_{\mu\nu} = \partial_{\nu}\,\textrm{ln}\sqrt{-\,G}=0\quad\quad\implies\quad\quad \tilde{\nabla}_{\mu}V^{\mu}=\partial_{\mu}V^{\mu}
\end{equation*}
For the Ricci tensor $\tilde{R}_{\mu\nu}$, this implies that
\begin{equation}\label{Eq:CompatibleF}
    \tilde{R}_{\mu\nu} = \partial_{\rho}\tilde{\Gamma}^{\rho}_{\mu\nu} - \tilde{\Gamma}^{\rho}_{\mu\lambda}\tilde{\Gamma}^{\lambda}_{\rho\nu}
\end{equation}
The equations of motion of an action-dependent field theory may be obtained via variational methods \cite{gaset2024herglotz}. However, the problem is a constrained variational calculation. More precisely, the $s^{\mu}$ are components of a $(d-1)$-form $s:=s^{\mu}d^{d-1}x_{\mu}$; the Herglotz principle states that the integral of $s$ over the boundary $\partial D$ of some domain $D \subset M$ must coincide with the value of the action of the solution. Symbolically, we have
\begin{equation}
    \int_D \mathcal{L}^H\circ j^1\phi = \int_{\partial D} s
\end{equation}
in which $\circ \; j^1\phi$ refers to the evaluation of $\mathcal{L}^H$ along the $1^{\textrm{st}}$ jet prolongation of the fields (c.f (\ref{Eq:MultisymplecticEOM2})). The Lagrange density, when evaluated on a given field configuration, is a top-form, and is thus closed. Locally, it follows that there exists a $(d-1)$-form, whose differential coincides with the Lagrangian density: this is precisely $s$. Consequently, we have $\mathcal{L}^H\circ j^1\phi = ds$; this constraint distinguishes non-conservative systems from their standard multisymplectic counterparts. Using a Lagrange multiplier, we introduce an extended action
\begin{equation}
    \widehat{S}=\int d^4x\;\big[(1-\lambda)\,\partial_\mu s^\mu+\lambda L^H\big]
\end{equation}
which may be varied freely. From variations of the action density, we ascertain the equation governing the evolution of $\lambda$
\begin{equation}
    \partial_{\mu}\lambda = -\,\lambda \frac{\partial L^H}{\partial s^{\mu}}
\end{equation}
In the present case, this extended action fails to reproduce the correct field equations when the background metric is dynamical. Even though $G$ is of fixed determinant, dynamical information is lost if this condition is not relaxed in the variational calculation. We thus reinstate the factor of $\sqrt{-G}$, and vary
\begin{equation}\label{Eq:HerglotzDensity}
    \mathcal{L}^H=\sqrt{-G}\left[ \frac{1}{2\kappa}\tilde{R}(G) - \frac{\kappa}{3}G_{\mu\nu}s^{\mu}s^{\nu}\right] d^4x
\end{equation}
subject to the modified Herglotz constraint $\partial_{\mu}(\sqrt{-G}\,s^{\mu})=\sqrt{-G}\,L^H$. Since we no longer require that $\sqrt{-G}=1$, the curvature $\tilde{R}$ is not of the simplified form (\ref{Eq:CompatibleF}); we must instead use the full expression
\begin{equation*}
    \tilde{R}_{\mu\nu} = \partial_{\rho}\tilde{\Gamma}^{\rho}_{\mu\nu} -\partial_{\mu}\tilde{\Gamma}^{\rho}_{\rho\nu} + \tilde{\Gamma}^{\rho}_{\rho\lambda}\tilde{\Gamma}^{\lambda}_{\mu\nu} - \tilde{\Gamma}^{\rho}_{\mu\lambda}\tilde{\Gamma}^{\lambda}_{\rho\nu}
\end{equation*}
Despite relaxing the condition $\sqrt{-G}=1$, not all metric variations are admissible. We denote an allowed variation $\delta^A G_{\mu\nu}$, noting that such objects are characterised by the requirement that $\delta^A\,\sqrt{-G}=0$. These restricted variations may be expressed in terms of an unconstrained $\delta G_{\mu\nu}$ according to
\begin{equation}\label{Eq:ConstrainedVar}
    \delta^A G_{\mu\nu} = \delta G_{\mu\nu} - \frac{1}{4}G^{\alpha\beta} G_{\mu\nu}\,\delta G_{\alpha\beta}
\end{equation}
It then follows that we may relate arbitrary and admissible variations of the action as
\begin{equation}\label{Eq:ConstrainedVar2}
    \frac{\delta S}{\delta^A G^{\mu\nu}} = \frac{\delta S}{\delta G^{\mu\nu}} - \frac{1}{4}G^{\mu\nu}G^{\alpha\beta}\frac{\delta S}{\delta G^{\alpha\beta}}
\end{equation}
Since the metric requires that the Herglotz condition be modified to $\partial_{\mu}(\sqrt{-G}\,s^{\mu})=\sqrt{-G}\,L^H$, we should consider an extended action of the form
\begin{equation}\label{Eq:ExtendedAction}
    \widehat{S}:=\int d^4x\;\sqrt{-G}\biggr[ (1-\lambda)\,\tilde{\nabla}_{\mu}s^{\mu} + \lambda L^H \biggr]
\end{equation}
Varying this action, making use of (\ref{Eq:ConstrainedVar2}, we find that the frictional field equations read
\begin{eqnarray}\label{Eq:FrictionalFieldEqns}
    \tilde{R}_{\mu\nu} - \frac{1}{4}\left(\tilde{R}+\frac{2\kappa^2}{9}G_{\alpha\beta}s^{\alpha}s^{\beta} - \frac{2\kappa}{3}\partial_{\alpha}s^{\alpha}\right)G_{\mu\nu} + \frac{2\kappa^2}{9}G_{\mu\alpha}G_{\nu\beta}s^{\alpha}s^{\beta} \label{Eq:A}\\
    -\frac{\kappa}{3}\left(G_{\mu\lambda}\tilde{\nabla}_{\nu}s^{\lambda} + G_{\lambda\nu}\tilde{\nabla}_{\mu}s^{\lambda}\right)=0\notag\\
    \tilde{\nabla}_{\mu}s^{\mu}=\partial_{\mu}s^{\mu} =L^H\label{Eq:B}
\end{eqnarray}
The steps leading to these frictional field equations resemble those used in the study of unimodular gravity (UG); indeed, the first term of (\ref{Eq:HerglotzLag2}) may be interpreted as a Lagrangian for just such a theory. However, it must be emphasised that our construction is \textit{not} simply UG, and differs in several crucial regards. In the unimodular approach, one \textit{chooses} to limit one's focus to metrics of unit determinant ab initio \cite{Alvarez_2005}. Our consideration of unimodular metrics, by contrast, was forced upon us after recognising that, of the ten independent degrees of freedom of $g_{\mu\nu}$, only those nine describing the conformal geometry of spacetime are indispensable for the dynamical evolution of observable quantities. Despite being a redundant degree of freedom, the conformal mode does hold status within the mathematical structure of the original Hilbert action. As such, the effect of the removal of this degree of freedom is not simply to pass from the Hilbert action with $g$ to the same action with $G$. Were this to be the case, our construction would indeed just yield UG. Instead, the removal of the conformal mode must be accompanied by a compensatory action dependence, whose presence is required in order to faithfully reproduce the original gravitational dynamics. This is precisely the content of the second term on the RHS of (\ref{Eq:HerglotzLag2}), and the feature which categorically separates our construction from UG.\\

Intuitively, at each point $x\in M$, the components $g_{\mu\nu}(x)$ may be considered to span a ten-dimensional space of symmetric matrices of Lorentzian signature $(-,+,+,+)$. In making the conformal decomposition, this factors locally into the product of an $\mathbb{R}_+$ and a nine-dimensional subspace, consisting of those Lorentzian metrics that have unit determinant. Quotienting by the orbits of the scaling symmetry eliminates precisely the $\mathbb{R}_+$, which we identify with a `radial' direction within the ten-dimensional space. The remaining metric components are then specified exclusively in terms of `angles'. This is similar in spirit to how we would write $\mathbb{R}^{10}\backslash \{0\}\cong \mathbb{R}_+\times S^9$; however, crucially, while $S^9$ is compact, the same is generally not true of the subspace $\{G_{\mu\nu}\;|\; \textrm{det}\,G=-1\,\}$, and so this statement is merely an analogy.\\

As verification of the consistency of our formalism, we should carry out the variation of the full Hilbert action, expressed in conformal variables. Using the relation between the action density and field velocities
\begin{equation}\label{Eq:SandG}
    s^{\mu}=\frac{3}{2\kappa}G^{\mu\nu}\partial_{\nu}\Phi
\end{equation}
the standard variational calculations which lead to the vacuum field equations should reproduce (\ref{Eq:FrictionalFieldEqns}). This is indeed the case, and is demonstrated explicitly in appendix \ref{App:Appendix1}. Note that it is important that (up to linear shifts, which can be absorbed by adding terms to the Herglotz Lagrangian) the action density be proportional to the gradient of a scalar. This requires that $s$ satisfy the integrability condition\footnote{Without this integrability condition, the solution space of the Herglotz theory is larger than that of GR.} $\partial_\mu s_\nu-\partial_\nu s_\mu=0$, of which we shall make further use when discussing the linearised limit.\\

As a final remark, we note that the construction presented thus far is valid in complete field-theoretic generality. Particular limiting cases have been studied in \cite{sloan2021new,sloan2021scale} and \cite{sloan2019scalar}, in the context of FLRW and (class-A) Bianchi cosmological models. It is a relatively straightforward exercise to make the appropriate simplifications for a spacetime that is homogeneous and/or isotropic. Taking $s^{\mu}=(S(t),0,0,0)$ (as must be the case to recover the particle constraint $\dot{S}=L^H$) the field equations (\ref{Eq:FrictionalFieldEqns}) are found to reduce to known results. Further, while we have focused exclusively on the vacuum case, our construction is perfectly amenable to the addition of (minimally-coupled) scalar fields, and so is fully compatible with, for example, models of inflationary cosmology.
\section{The Weak-Field Limit}\label{Sec:WeakFieldLimit}
In the presence of weak gravitational fields, it is illustrative to study the linearised limit of the gravitational field equations. We write the spacetime metric as $g_{\mu\nu}=\eta_{\mu\nu}+h_{\mu\nu}$, in which $h_{\mu\nu}$ describes a small deviation from a flat Minkowski space. The field equations are then expanded to first order in $h_{\mu\nu}$, and the resulting theory found to describe the propagation of a symmetric rank-two covariant tensor field on a flat background. More generally, taking the linearised limit around any background spacetime, the perturbations are found to behave covariantly under the isometry group of that background. Having shown that GR admits a scale-free description, it will be of interest to examine the linear limit of this theory. To this end, we introduce the following weak-field approximations
\begin{equation}\label{Eq:WeakField}
    G_{\mu\nu}=\eta_{\mu\nu}+H_{\mu\nu} \hspace{2.8cm} G^{\mu\nu}=\eta^{\mu\nu}-H^{\mu\nu}
\end{equation}
in which indices are raised and lowered with $\eta^{\mu\nu}$ and $\eta_{\mu\nu}$. Note that, in addition to being symmetric, our perturbation must also be traceless, since
\begin{equation*}
    \textrm{det}\left(\eta_{\mu\nu}+H_{\mu\nu}\right) = -\left(1+ H\right) + \mathcal{O}(H^2),\hspace{2cm} H:=\eta^{\mu\nu}H_{\mu\nu}
\end{equation*}
Action dependence of the Herglotz Lagrangian requires that we also expand the action density; for a flat background, the zeroth-order value of $s^{\mu}$ should be zero. Thus, to first order, we write $s^{\mu}=\sigma^{\mu}$. It is straightforward to expand (\ref{Eq:FrictionalFieldEqns}) to first order in the perturbations; we find that
\begin{equation}\label{Eq:FirstOrder}
    \begin{split}
        \frac{1}{2}\left(\partial_{\rho}\partial_{\mu} H^{\rho}_{\;\nu} + \partial_{\rho}\partial_{\nu} H^{\rho}_{\;\mu} - \square H_{\mu\nu} \right) - \frac{1}{2}\left(\partial_{\alpha}\partial_{\beta}H^{\alpha\beta} - \frac{2\kappa}{3} \partial_{\alpha}\sigma^{\alpha}\right) \eta_{\mu\nu} - \frac{\kappa}{3}\left(\eta_{\mu\lambda}\partial_{\nu}\sigma^{\lambda} +\eta_{\lambda\nu}\partial_{\mu}\sigma^{\lambda}\right)=0
    \end{split}
\end{equation}
Note that the first-order expansion of $\tilde{R}_{\mu\nu}$ does not contain the usual factor of $\partial_{\mu}\partial_{\nu} H$, as $H:=\eta^{\mu\nu}H_{\mu\nu}$ is vanishing as a result of the unimodularity condition. At present, our theory is invariant under a large class of gauge transformations; in particular, infinitesimal diffeomorphisms generated by vector fields $\xi^{\mu}(x)$, such that $\partial_{\mu}\xi^{\mu}=0$, do not alter the physical content of our description. Ordinarily, one uses part of this freedom to impose the harmonic gauge, which may be expressed as the requirement that $g^{\mu\nu}\Gamma_{\mu\nu}^{\lambda}=0$. Rewriting this condition using the results of the previous section, we have
\begin{equation}\label{Eq:HarmonicGauge}
    G^{\mu\nu}\tilde{\Gamma}_{\mu\nu}^{\lambda} - \frac{2\kappa}{3}s^{\lambda}=0
\end{equation}
Expanding to first order, we lower the $\lambda$ index with the Minkowski metric, obtaining
\begin{equation}\label{Eq:GaugeChoice}
    \partial_{\rho}H^{\rho}_{\;\mu} - \frac{2\kappa}{3}\eta_{\mu\lambda}\sigma^{\lambda}=0
\end{equation}
A priori, it seems we have fixed more degrees of freedom than those afforded to us by unimodular diffeomorphisms. A vector field $\xi^{\mu}$ provides four independent choices; however, the divergence-free condition reduces this to just three. The expression (\ref{Eq:GaugeChoice}) contains four components, and so appears excessively restrictive. The reconciliation lies in that not all four components of (\ref{Eq:GaugeChoice}) are independent. Taking the derivative $\partial^{\mu}$ of the gauge-fixing condition, we obtain a term of the form $\partial_{\mu}\sigma^{\mu}$; this is subject to the Herglotz constraint, and so must coincide with the first-order expansion of $L^H$. Consequently, (\ref{Eq:GaugeChoice}) comprises only three independent degrees of freedom.\\

On-shell, $s^{\mu}$ is a coordinate-dependent variable, and so under an infinitesimal shift $x^{\mu} \rightarrow x^{\mu} + \xi^{\mu}(x)$, is subject to a transformation. To first order, however, this transformation is proportional to the zeroth-order value of $s^{\mu}$, which is zero. Hence, we we find that, under the infinitesimal coordinate transformation, the gauge condition (\ref{Eq:GaugeChoice}) transforms as
\begin{equation*}
    \delta_{\xi}\left(\partial_{\rho}H^{\rho}_{\;\mu} - \frac{2\kappa}{3}\eta_{\mu\lambda}\sigma^{\lambda} \right) = \square \,\xi_{\mu} \quad\quad\textrm{with}\;\,\partial_{\mu}\xi^{\mu}=0
\end{equation*}
It thus follows that we may use a unimodular diffeomorphism to pass to this gauge. Such a choice reduces (\ref{Eq:FirstOrder}) to the standard free wave equation $\square H_{\mu\nu}=0$. While this is not unexpected, let us examine the consequences of proposing a plane wave solution of the form
\begin{equation}
    H_{\mu\nu}=C_{\mu\nu} e^{ik\cdot x}
\end{equation}
with $C_{\mu\nu}$ symmetric and traceless. Non-trivial solutions of the wave equation require that $\eta_{\mu\nu}k^{\mu}k^{\nu}=0$, as usual. Imposing the harmonic gauge condition (\ref{Eq:GaugeChoice}), we have
\begin{equation}\label{Eq:OscillatorySigma}
    ik_{\rho}C^{\rho}_{\mu} e^{ik\cdot x} = \frac{2\kappa}{3}\eta_{\mu\rho}\sigma^{\rho}\quad\quad\implies\quad\quad \sigma^{\mu}=\frac{3i}{2\kappa}C^{\mu\nu}k_{\nu}e^{ik\cdot x}
\end{equation}
And so the frictional degrees of freedom must also behave in an oscillatory manner. Note that the algebraic structure has changed somewhat with respect to the standard treatment of the linearised field equations. In particular, we would typically introduce the trace-reversed perturbation, and propose a plane-wave ansatz for this quantity. The harmonic gauge condition would then impose that the polarisation tensor be orthogonal to the wave vector. In this symmetry-reduced setup, however, the perturbation $H_{\mu\nu}$ is required to be traceless ab initio, as a result of the requirement that $\text{det}\,G=-1$. As such, we cannot construct a trace-reversed $H_{\mu\nu}$, and the harmonic gauge condition instead requires that the action density $\sigma^\mu$ have oscillatory behaviour, with polarisation proportional to $C_{\mu\nu}k^\nu$. Note that this does not a priori forbid $C_{\mu\nu}k^\nu=0$, which would make $\sigma^\mu=0$, but transversality is not directly \textit{implied} by the harmonic gauge condition.\\

We do, however, have two more opportunities to further restrict the polarisation tensor. The first of these is the familiar residual gauge symmetry, afforded by those coordinate transformations whose generator $\zeta^\mu$ is harmonic. In our case, we must also ensure that $\partial_\mu\zeta^\mu=0$, else the diffeomorphism is not unimodular. The second way in which we further restrict the components of $C_{\mu\nu}$ is the Herglotz condition. In order for our action-dependent theory to be consistent, we require that $\partial_\mu s^\mu=L^H$ be satisfied order-by-order. The zeroth-order expansion is trivial; however, at first order, we find that
\begin{equation}
    \partial_{\mu}\sigma^{\mu} \stackrel{!}{=} \frac{1}{2\kappa}\partial_{\mu}\partial_{\nu}H^{\mu\nu} \quad\quad\implies \quad\quad -\,\frac{3}{2\kappa}C^{\mu\nu}k_{\mu}k_{\nu}e^{ik\cdot x} = -\,\frac{1}{2\kappa}C^{\mu\nu}k_{\mu}k_{\nu} e^{ik\cdot x}
\end{equation}
From this, consistency gives us the single constraint $C^{\mu\nu}k_{\mu}k_{\nu}=0$.\\

Under the residual transformation $x^\mu\rightarrow x^\mu + \zeta^\mu$, with $\square\zeta^\mu=0$ and $\partial_\mu\zeta^\mu=0$, the condition (\ref{Eq:GaugeChoice}) is left unchanged, while $H_{\mu\nu}\rightarrow H_{\mu\nu}+\partial_\mu\zeta_\nu + \partial_\nu\zeta_\mu$. Since $\square\zeta_\mu=0$ is just a free wave equation for $\zeta_\mu$ we take a plane wave solution of the form $\zeta_\mu=v_\mu e^{ik\cdot x}$. Having specified this solution for $\zeta_\mu$, we have now completely fixed the gauge, and there is no residual freedom. The polarisation tensor transforms as $C_{\mu\nu}\rightarrow C_{\mu\nu}+i\left(k_\mu v_\nu + k_\nu v_\mu\right)$. However, the freedom to choose $v_\mu$ \textit{still} does not let us impose that $C_{\mu\nu}k^\nu=0$, since the null property of $k^\mu$, together with the unimodularity of the transformation generated by $\zeta$ (which requires that $k^\mu v_\mu=0$) implies that $k^\nu\left[C_{\mu\nu}+i\left(k_\mu v_\nu + k_\nu  v_\mu\right)\right]=C_{\mu\nu}k^\nu$, and so our residual gauge transformation cannot alter the value of $C_{\mu\nu}k^\nu$.\\

We therefore have a slightly alarming situation, in which, despite knowing that gravitational radiation possesses only two physical, propagating degrees of freedom, each of which is orthogonal to the wave vector, the mathematics simply does not seem to require this. Naturally, there is a resolution: a subtlety we have overlooked in the relation (\ref{Eq:SandG}) between $s^\mu$ and the conformal mode $\Phi$. Specifically, since we have expanded to first order, (\ref{Eq:SandG}) requires that
\begin{equation*}
    \sigma^\mu=\frac{3}{2\kappa}\partial^\mu\Phi^{(1)}
\end{equation*}
in which $\Phi^{(1)}$ denotes the first-order expansion of $\Phi$. Thus, in order for $\sigma$ to represent a valid physical configuration coming from a contact reduction, we see that it must be written as the gradient of a scalar. As discussed at the end of section (\ref{Sec:Metric}), this requires that $\sigma$ satisfy the relatively strict integrability condition $\partial_\mu\sigma_\nu-\partial_\nu\sigma_\mu=0$. If we now substitute our result (\ref{Eq:OscillatorySigma}) into this constraint, we find that
\begin{equation*}
    ik_{\mu }\left(\frac{3i}{2\kappa }C_{\nu \rho }k^{\rho }\right)e^{ik\cdot x}-ik_{\nu }\left(\frac{3i}{2\kappa }C_{\mu \rho }k^{\rho }\right)e^{ik\cdot x}=0 \quad\quad\implies\quad\quad k_\mu(C_{\nu\rho}k^\rho) = k_\nu(C_{\mu\rho}k^\rho)
\end{equation*}
The only way for this condition to be satisfied for all $\mu,\nu$ is if $C_{\mu\nu}k^\nu = ak_\mu$, for some $a\in\mathbb{R}$. With this, we can integrate the above relationship between $\sigma$ and $\Phi^{(1)}$, to deduce that $\Phi^{(1)}=ae^{ik\cdot x}$. This reveals that those configurations for which $a\neq0$ (which, in turn, implies that $C_{\mu\nu}k^\nu\neq0$, giving rise to the problematic longitudinal modes) coincide precisely with a coordinate-dependent rescaling of the conformal factor.\\

In the standard setting, we have the full diffeomorphism group $\text{Diff}(M)$ at our disposal. When we write the complete spacetime metric as $g_{\mu\nu}=e^\Phi G_{\mu\nu}$, and expand to first order $g_{\mu\nu}=\eta_{\mu\nu}+h_{\mu\nu}$, we find that the full perturbation $h_{\mu\nu}$ splits into a shape piece, which is exactly our $H_{\mu\nu}$, together with a pure scale piece $\Phi^{(1)}\eta_{\mu\nu}$. In this framework, when a gravitational wave has a non-zero conformal scale component ($\Phi^{(1)}\neq0$), it is always possible to perform a non-volume-preserving diffeomorphism, so as to eliminate the scale component, and force the oscillations into $H_{\mu\nu}$, in a manner such that they are transverse to the wave vector. Upon making the contact reduction, we have eliminated $\Phi$ from our ontology, and with it, the possibility of carrying out non-volume-preserving diffeomorphisms. Because of this, we have lost the freedom to change the local volume via gauge, and if we begin with a configuration with $C_{\mu\nu}k^\nu\neq0$, there is no allowed transformation that we can make that will change this.\\

The reconciliation is then that we have two distinct cases. The simplest and most physically transparent is that the configuration does not have any kind of scaling that we would have to eliminate with a non-volume-preserving diffeomorphism. This is precisely the case $a=0$. Here, $C_{\mu\nu}k^\nu=0$, and the wave manifestly possesses two transverse propagating degrees of freedom. The alternative is that $a\neq0$, so that $C_{\mu\nu}k^\nu\neq0$. This is still a perfectly valid solution, but requires more care with its physical interpretation. Here, the fact that $a\neq0$ means that, in the full theory, the perturbation looks like $h_{\mu\nu}=H_{\mu\nu} + \Phi^{(1)}\eta_{\mu\nu}$, so that the transformation taking us to the transverse-traceless gauge must necessarily alter the volume, in order to eliminate the $\Phi^{(1)}\eta_{\mu\nu}$ piece, which is manifestly non-transverse. In the reduced description, the concept of local scaling has been completely eliminated; all we have are volume-preserving diffeomorphisms and friction. The action density is forced to oscillate in a direction perpendicular\footnote{This follows from the fact that $\sigma^\mu\propto C^{\mu\nu}k_\nu e^{ik\cdot x}$, and $k^\mu(C_{\mu\nu}k^\nu)=0$.} to $k^\mu$, in order to compensate what is, in the original description, an oscillating volume element.\\

The above series of arguments does \textit{not} imply the presence of new degrees of freedom. Indeed, a simple counting reveals that the traceless symmetric polarisation $C_{\mu\nu}$ has nine degrees of freedom, of which six are removed by volume-preserving diffeomorphisms, and a further one is eliminated through the constraint $C_{\mu\nu}k^\mu k^\nu=0$. This leaves the expected value of two. Instead, we should understand that the elimination of the conformal mode changes the way in which the constraints are distributed between gauge degrees of freedom and dynamics. In particular, in those cases for which $C_{\mu\nu}k^\nu\neq0$, we saw that the field equations themselves imposed that $\sigma^\mu$ acquired oscillatory characteristics.\\

Having studied in detail the first-order expansion, we now turn to an analysis of the field equations at second order. Here, the frictional degrees of freedom do not decouple from the wave dynamics, and our theory will be manifestly non-conservative. We write the second-order expansions as follows
\begin{equation}\label{Eq:2ndOrder}
    G_{\mu\nu} = \eta_{\mu\nu}+H_{\mu\nu} + \bar{H}_{\mu\nu}\hspace{2.5cm} s^{\mu}=\sigma^{\mu}+\bar{\sigma}^{\mu} 
\end{equation}
We will express the expansion of the field equations (\ref{Eq:FrictionalFieldEqns}) schematically as
\begin{equation}\label{Eq:Schematic}
    \Gamma_{\mu\nu}^{(1)}[H,\sigma] + \Gamma_{\mu\nu}^{(2)}[\bar{H},\bar{\sigma}]=0
\end{equation}
in which $\Gamma_{\mu\nu}^{(1)}[\bar{H},\bar{\sigma}]$ refers to those terms linear in the second-order parameters, and $\Gamma_{\mu\nu}^{(2)}[H,\sigma]$ contains terms quadratic in $H_{\mu\nu}$ and $\sigma^{\mu}$. After some work, we find that $\Gamma_{\mu\nu}^{(1)}[\bar{H},\bar{\sigma}]$ is given by
\begin{equation}\label{Eq:2ndOrderExpansion1}
    \begin{split}
        \Gamma_{\mu\nu}^{(1)}[\bar{H},\bar{\sigma}] = \frac{1}{2}\left(\partial_{\rho}\partial_{\mu} \bar{H}^{\rho}_{\;\nu} + \partial_{\rho}\partial_{\nu} \bar{H}^{\rho}_{\;\mu} - \partial_{\mu}\partial_{\nu} \bar{H} - \square \bar{H}_{\mu\nu} \right) - \frac{1}{4}\left(\partial_{\alpha}\partial_{\beta}\bar{H}^{\alpha\beta} - \square\bar{H} - \frac{2\kappa}{3} \partial_{\alpha}\bar{\sigma}^{\alpha}\right) \eta_{\mu\nu}\\
        - \frac{\kappa}{3}\left(\eta_{\mu\lambda}\partial_{\nu}\bar{\sigma}^{\lambda} +\eta_{\lambda\nu}\partial_{\mu}\bar{\sigma}^{\lambda}\right)
    \end{split}
\end{equation}
Note that, unlike the case of the first-order expansion, here $\bar{H}\neq 0$. The second-order piece $\Gamma_{\mu\nu}^{(2)}[H,\sigma]$ contributes the following terms
\begin{equation}\label{Eq:2ndOrderExpansion2}
    \begin{split}
        \Gamma_{\mu\nu}^{(2)}[H,\sigma] = \tilde{R}^{(2)}_{\mu\nu} - \frac{1}{4}\left(\tilde{R}^{(2)} -\frac{1}{2}H^{\alpha\beta}\left(2\,\partial_{\rho}\partial_{\alpha} H^{\rho}_{\beta} - \square H_{\alpha\beta}\right) + \frac{2\kappa^2}{9}\eta_{\alpha\beta}\sigma^{\alpha}\sigma^{\beta}\right)\eta_{\mu\nu}\\
        -\frac{\kappa}{3}\left(\sigma^{\rho}\partial_{\rho}H_{\mu\nu} + H_{\mu\rho}\partial_{\nu}\sigma^{\rho}+H_{\rho\nu}\partial_{\mu}\sigma^{\rho} \right) + \frac{2\kappa^2}{9}\eta_{\mu\alpha}\eta_{\nu\beta}\sigma^{\alpha}\sigma^{\beta}
    \end{split}
\end{equation}
in which $\tilde{R}^{(2)}_{\mu\nu}$ denotes those terms in the expansion of the Ricci tensor which contain terms schematically of the form $\partial H\,\partial H$ and $H\partial\partial H$, and $\tilde{R}^{(2)}:=\eta^{\mu\nu}\tilde{R}^{(2)}_{\mu\nu}$. In the above expansion, we have also omitted a term proportional to the gauge condition (\ref{Eq:GaugeChoice}). The full field equations expanded to second order are simply the sum of (\ref{Eq:2ndOrderExpansion1}) and (\ref{Eq:2ndOrderExpansion2}); the second-order gauge condition may be conveniently expressed as
\begin{equation}\label{Eq:GaugeChoice2}
    \partial_{\rho} \bar{H}^{\rho}_{\mu} -\frac{2\kappa}{3}\eta_{\mu\rho}\bar{\sigma}^{\rho} = \frac{1}{2}\partial_{\mu}\bar{H} + H^{\alpha\beta}\partial_{\beta} H_{\mu\alpha} - \frac{1}{2}H^{\alpha\beta}\partial_{\mu}H_{\alpha\beta} + H^{\alpha}_{\mu}\partial_{\beta} H^{\beta}_{\alpha}
\end{equation}
With this, the final second-order expansion becomes
\begin{equation}\label{Eq:2ndOrderFieldEqns}
    \begin{split}
        \square \bar{H}_{\mu\nu} - \frac{1}{4}\square \bar{H}\eta_{\mu\nu} + \frac{2\kappa}{3}\left(\sigma^{\lambda}\partial_{\lambda}H_{\mu\nu} + H_{\mu\lambda}\partial_{\nu}\sigma^{\lambda} + H_{\lambda\nu}\partial_{\mu}\sigma^{\lambda}\right) = Q_{\mu\nu} + C\eta_{\mu\nu} \\
        + \frac{4\kappa^2}{9}\left( \eta_{\mu\alpha}\eta_{\beta\nu}-\frac{1}{4}\eta_{\mu\nu}\eta_{\alpha\beta}\right)\sigma^{\alpha}\sigma^{\beta}
    \end{split}
\end{equation}
Here, we have collected all terms of the form $\partial H\,\partial H$ and $H\partial\partial H$ into the object $Q_{\mu\nu}$, which includes the second-order expansion $\tilde{R}^{(2)}_{\mu\nu}$. Similarly, terms quadratic in $H$ which have no free indices have been grouped into the object we have called $C$.\\

Physically, the second-order expansion describes the self-interaction of the propagating modes. The first two terms on the left-hand side of (\ref{Eq:2ndOrderFieldEqns}) correspond to the standard kinetic structure, modified by the unimodularity condition. The remaining terms, involving $\sigma^\mu$, reflect a non-trivial coupling between the wave degrees of freedom and the action-density sector, signaling the intrinsically non-conservative nature of the theory. In contrast with the standard framework of GR, where the second-order dynamics may be interpreted in terms of an effective conserved energy-momentum tensor for gravitational waves, here the backreaction is not governed by a conserved quantity. Instead, energy is exchanged between the geometric perturbations and the additional degrees of freedom encoded in $s^\mu$. This provides a dynamical realisation of the fact that, following the elimination of the conformal mode, the constraints of the theory are redistributed between gauge symmetry and dynamics.
\section{Discussion and Conclusions}\label{Sec:Conclusion}
In this work, we have analysed a reformulation of classical General Relativity, in which the conformal degree of freedom has been identified as redundant. The evolution of the remaining dynamical variables may be faithfully reproduced without reference to this ontologically-inaccessible parameter. While dynamically equivalent to standard General Relativity, our contact-reduced theory exhibits a reduced gauge symmetry, being invariant only under volume-preserving diffeomorphisms, and requires a qualitative reinterpretation, as an intrinsically non-conservative description, which is made manifest by an action-dependent Lagrangian.\\

While the first-order formalism provided the correct geometrical setting in which to make the symmetry reduction, the slightly unwieldy nature of the frame field algebra led us to recast our results in terms of a second-order metric theory. Such a change in description was particularly appropriate for our subsequent study of the weak-field limit, in which we linearised the field equations around a flat Minkowski background. It was shown that, at the level of linear perturbations, the theory propagates the expected two physical degrees of freedom. However, the reduced gauge symmetry prevents the imposition of full transversality, and the elimination of non-physical components is achieved through a combination of residual gauge freedom and dynamical constraints.\\

At second order, the differences became more pronounced; we observed that the backreaction of gravitational waves was no longer governed by an effective conserved energy-momentum tensor, but instead involved a non-trivial coupling to the action-density sector. This led to a departure from the standard interpretation of gravitational wave energy, and reflects the fundamentally non-conservative nature of the theory.\\

One particularly interesting aspect of our construction lies in the fact that we are able to reproduce the complete (observable) dynamical content of General Relativity, as captured by the Hilbert action, without need to reference the conformal factor $\phi$. Indeed, this degree of freedom is frequently assigned the physical interpretation of a volume/scale parameter. In such cases, this interpretation becomes highly non-physical when we approach the initial singularity, where the standard mathematical framework becomes pathological. It was shown in \cite{sloan2019scalar} that for the simple case of a matter-sourced FLRW model, while the conventional description presents non-predictive behaviour at the initial singularity, the contact-reduced model suffered no such pathologies. Indeed, it was found that the solution could be continued in a perfectly predictive manner, and that traversing the singularity was associated with a change in the orientation of the spacetime manifold. The results of the present work constitute a complete field-theoretic generalisation of the contact reduction of classical General Relativity, and so naturally subsume the content of \cite{sloan2019scalar}. It would be of great interest, therefore, to examine whether techniques similar to those used for the particular case of homogeneous cosmologies may be utilised to draw more general conclusions about the nature of General Relativity in the vicinity of singularities, in which spacetime volumes collapse to zero.\\

In \cite{Bell:2025ooq}, it was found that the processes of contact reduction and phase space restriction are commutative. This is not surprising; however, it raises the interesting possibility of performing a $3+1$ decomposition of the contact-reduced Lagrangian (\ref{Eq:HerglotzLag2}), passing to a contact Hamiltonian description. It would be illustrative to examine whether any components of the constraint analysis are simplified by working directly with the reduced ontology, which makes no reference to the conformal factor. Indeed, more broadly, our work highlights how dynamically equivalent formulations of a single classical theory can offer novel insight, and in some cases/regimes - as exemplified by our discussion of singularities - be a more suitable description than the standard framework.
\appendix
\section{Frictional Field Equations}\label{App:Appendix1}
Upon excising the conformal factor from our ontology, we obtained a first-order action-dependent Lagrangian of the form
\begin{equation}\label{Eq:App1HerglotzLagrangian}
    \begin{split}
        L^H = \frac{1}{2\kappa} \tilde{e}^{\mu}_{\;I}\tilde{e}^{\nu}_{\;J}\tilde{R}_{\mu\nu}^{\quad IJ} - \frac{\kappa}{3} \, \tilde{e}_{\mu}^{\;K} \tilde{e}_{\nu K} s^{\mu}s^{\nu} + \frac{1}{3} s^{\mu} \tilde{e}_{\mu}^{\;[I}\tilde{e}_{\lambda}^{\;J]}\tilde{\mathcal{D}}_{\sigma}(\tilde{e}^{\sigma}_{\;I}\tilde{e}^{\lambda}_{\;J})\\
        - \frac{1}{12\kappa}\,\tilde{e}_{\mu}^{\;[I}\tilde{e}_{\rho}^{\;J]}\,\tilde{e}^{\mu[A}\tilde{e}_{\lambda}^{\;B]}\,\tilde{\mathcal{D}}_{\sigma}\left(\tilde{e}^{\sigma}_{\;I}\tilde{e}^{\rho}_{\;J}\right)\tilde{\mathcal{D}}_{\alpha}\left(\tilde{e}^{\alpha}_{\;A}\tilde{e}^{\lambda}_{\;B}\right)
    \end{split}
\end{equation}
We shall demonstrate that, as claimed in the discussion surrounding (\ref{Eq:HerglotzLagrangian}), the terms containing covariant derivatives of the frame fields vanish on-shell. Having done so, we shall pass to the second-order metric formalism, proving that variation of the multisymplectic action (\ref{Eq:EHAction}), expressed in conformal variables, does indeed reproduce the field equations (\ref{Eq:FrictionalFieldEqns}).\\

As discussed in the main text, action-dependent theories may treated as unconstrained variational problems, provided a Lagrange multiplier is inserted to enforce the Herglotz condition. Since we consider only variations of $\tilde{\omega}^{IJ}$, and not of the frame field, we need not concern ourselves with any of the subtleties related to the unit-determinant condition (see discussion surrounding (\ref{Eq:HerglotzDensity})). It therefore suffices to consider 
\begin{equation*}
    \widehat{S}:=\int d^4x\biggr[ (1-\lambda)\,\partial_{\mu}s^{\mu} + \lambda L^H \biggr]
\end{equation*}
An expression for the derivative $\partial_{\mu}\lambda$ of the Lagrange multiplier will be required, whenever partial integration is carried out; varying $\widehat{S}$ with respect to $s^{\mu}$, we find that
\begin{equation*}
    \begin{split}
        \delta\widehat{S}&= \int d^4x \;\biggr[(1-\lambda)\,\delta(\partial_{\mu}s^{\mu}) + \frac{\lambda}{3}\left(-2\kappa \tilde{e}_{\mu}^{\;K} \tilde{e}_{\nu K}s^{\nu} + \tilde{e}_{\mu}^{\;[I}\tilde{e}_{\lambda}^{\;J]}\tilde{\mathcal{D}}_{\sigma}(\tilde{e}^{\sigma}_{\;I}\tilde{e}^{\lambda}_{\;J})\right)\delta s^{\mu} \biggr]\\
        &= \int d^4x \;\biggr[\partial_{\mu}\lambda - \frac{\lambda}{3}\left(2\kappa \tilde{e}_{\mu}^{\;K} \tilde{e}_{\nu K}s^{\nu} - \tilde{e}_{\mu}^{\;[I}\tilde{e}_{\lambda}^{\;J]}\tilde{\mathcal{D}}_{\sigma}(\tilde{e}^{\sigma}_{\;I}\tilde{e}^{\lambda}_{\;J})\right) \biggr]\,\delta s^{\mu}
    \end{split}
\end{equation*}
from which it follows that
\begin{equation}\label{Eq:App1Lagrange}
    \partial_{\mu}\lambda = \frac{\lambda}{3}\left(2\kappa \tilde{e}_{\mu}^{\;K} \tilde{e}_{\nu K}s^{\nu} - \tilde{e}_{\mu}^{\;[I}\tilde{e}_{\lambda}^{\;J]}\tilde{\mathcal{D}}_{\sigma}(\tilde{e}^{\sigma}_{\;I}\tilde{e}^{\lambda}_{\;J})\right)
\end{equation}
We may now consider varying $\widehat{S}$ with respect to $\tilde{\omega}^{IJ}$. The second term on the right-hand side of (\ref{Eq:App1HerglotzLagrangian}) contains no reference to the spin connection; it will be of benefit to expand each of the covariant derivative terms, making explicit their dependence on $\tilde{\omega}^{IJ}$
\begin{equation}
    \begin{split}
        L^H \;\supset\;\; &\frac{1}{2\kappa} \tilde{e}^{\mu}_{\;I}\tilde{e}^{\nu}_{\;J}\tilde{R}_{\mu\nu}^{\quad IJ} - \frac{1}{3} s^{\mu} \tilde{e}_{\mu [I}\tilde{e}_{\rho J]}\left( \tilde{\omega}_{\sigma}^{\;\;KI} \, \tilde{e}^{\sigma}_{\;K} \tilde{e}^{\rho J} + \tilde{\omega}_{\sigma}^{\;\;KJ} \, \tilde{e}^{\sigma I}\tilde{e}^{\rho}_{\;K}\right) \\
        &+ \frac{1}{12\kappa}\,\tilde{e}_{\mu [I}\tilde{e}_{\rho J]}\,\tilde{e}^{\mu[A}\tilde{e}_{\lambda}^{\;B]}\left(\tilde{\omega}_{\sigma}^{\;\;KI}\,\tilde{e}^{\sigma}_{\;K}\tilde{e}^{\rho J} + \tilde{\omega}_{\sigma}^{\;\;KJ} \, \tilde{e}^{\sigma I}\tilde{e}^{\rho}_{\;K}\right) \tilde{\mathcal{D}}_{\alpha}(\tilde{e}^{\alpha}_{\;A}\tilde{e}^{\lambda}_{\;B})
    \end{split}
\end{equation}
Note that, in expanding the covariant derivatives, we have intentionally omitted any terms such as $\partial_{\sigma}(\tilde{e}^{\sigma}_{\;I}\tilde{e}^{\lambda}_{\;J})$, as these do not contribute to our calculation. We have also swapped upper and lower $I$ and $J$ indices, as compared to (\ref{Eq:App1HerglotzLagrangian}). Not forgetting the overall multiplicative factor of $\lambda$, variations of $\widehat{S}$ with respect to $\tilde{\omega}_{\mu}^{\;\;IJ}$ yield
\begin{equation*}
    \begin{split}
        \delta\widehat{S} = \int d^4x\,\lambda&\left[\frac{1}{2\kappa}\tilde{e}^{[\mu}_{\;I}\tilde{e}^{\nu]}_{\;J}\tilde{\mathcal{D}}_{\mu}\delta \tilde{\omega}_{\nu}^{\;\;IJ} -\frac{1}{3} s^{\mu} \tilde{e}_{\mu [I}\tilde{e}_{\rho J]}\left(  \tilde{e}^{\sigma}_{\;K} \tilde{e}^{\rho J}\delta\tilde{\omega}_{\sigma}^{\;\;KI} +  \tilde{e}^{\sigma I}\tilde{e}^{\rho}_{\;K} \delta \tilde{\omega}_{\sigma}^{\;\;KJ} \right)\right.\\
        & + \left.\frac{1}{6\kappa} \tilde{e}_{\mu [I}\tilde{e}_{\rho J]}\,\tilde{e}^{\mu[A}\tilde{e}_{\lambda}^{\;B]}\left(\tilde{e}^{\sigma}_{\;K}\tilde{e}^{\rho J}\delta\tilde{\omega}_{\sigma}^{\;\;KI} + \tilde{e}^{\sigma I}\tilde{e}^{\rho}_{\;K}\delta\tilde{\omega}_{\sigma}^{\;\;KJ} \right) \tilde{\mathcal{D}}_{\alpha}(\tilde{e}^{\alpha}_{\;A}\tilde{e}^{\lambda}_{\;B})\right]\\
        = \int d^4x\,\lambda&\left[\frac{1}{2\kappa}\tilde{e}^{[\mu}_{\;I}\tilde{e}^{\nu]}_{\;J}\tilde{\mathcal{D}}_{\mu}\delta \tilde{\omega}_{\nu}^{\;\;IJ} -\frac{1}{3}\tilde{e}_{\mu [I}\tilde{e}_{\rho J]}\left(s^{\mu}-\frac{1}{2\kappa}\tilde{e}^{\mu[A}\tilde{e}_{\lambda}^{\;B]}\tilde{\mathcal{D}}_{\alpha}(\tilde{e}^{\alpha}_{\;A}\tilde{e}^{\lambda}_{\;B})\right)\left( \tilde{e}^{\sigma}_{\;K} \tilde{e}^{\rho J}\delta\tilde{\omega}_{\sigma}^{\;\;KI} +  \tilde{e}^{\sigma I}\tilde{e}^{\rho}_{\;K} \delta \tilde{\omega}_{\sigma}^{\;\;KJ}\right)\right]
    \end{split}
\end{equation*}
We recognise the bracketed term containing $s^{\mu}$ as a rearrangement of (\ref{Eq:ActionDensity}). In order to simplify the action further, we observe that
\begin{equation*}
    \tilde{e}_{\mu [I}\tilde{e}_{\rho J]}\left( \tilde{e}^{\sigma}_{\;K} \tilde{e}^{\rho J}\delta\tilde{\omega}_{\sigma}^{\;\;KI} +  \tilde{e}^{\sigma I}\tilde{e}^{\rho}_{\;K} \delta \tilde{\omega}_{\sigma}^{\;\;KJ}\right)= 2\,\tilde{e}_{\mu I}\tilde{e}^{\sigma}_{\;K} \delta\tilde{\omega}_{\sigma}^{\;\;KI}
\end{equation*}
Inserting this into our expression for $\delta\widehat{S}$, we have
\begin{equation*}
    \delta\widehat{S}=\int d^4x\,\lambda \left[\frac{1}{2\kappa}\tilde{e}^{[\mu}_{\;I}\tilde{e}^{\nu]}_{\;J}\tilde{\mathcal{D}}_{\mu}\delta \tilde{\omega}_{\nu}^{\;\;IJ} -\frac{2}{3}\tilde{e}_{\mu I}\tilde{e}^{\sigma}_{\;K} \left(s^{\mu}-\frac{1}{2\kappa}\tilde{e}^{\mu[A}\tilde{e}_{\lambda}^{\;B]}\tilde{\mathcal{D}}_{\alpha}(\tilde{e}^{\alpha}_{\;A}\tilde{e}^{\lambda}_{\;B})\right)\delta\tilde{\omega}_{\sigma}^{\;\;KI}\right]
\end{equation*}
This expression is now of a form that may be integrated by parts; in particular, we integrate the first term, making use of (\ref{Eq:App1Lagrange}) to rewrite $\partial_{\mu}\lambda$ as something proportional to $\lambda$
\begin{equation*}
     \begin{split}   
        \delta\widehat{S}=-\int d^4x\,\lambda \left[\frac{1}{2\kappa}\left( \frac{1}{3}\tilde{e}^{[\mu}_{\;I}\tilde{e}^{\nu]}_{\;J}\left(2\kappa \tilde{e}_{\mu}^{\;K} \tilde{e}_{\lambda K}s^{\lambda} - \tilde{e}_{\mu}^{\;[A}\tilde{e}_{\lambda}^{\;B]}\tilde{\mathcal{D}}_{\sigma}(\tilde{e}^{\sigma}_{\;A}\tilde{e}^{\lambda}_{\;B})\right) + \tilde{\mathcal{D}}_{\mu}(\tilde{e}^{[\mu}_{\;I}\tilde{e}^{\nu]}_{\;J})\right)\delta\tilde{\omega}_{\nu}^{\;\;IJ}\right.\\
        \left. + \frac{2}{3}\tilde{e}_{\mu I}\tilde{e}^{\sigma}_{\;K} \left(s^{\mu}-\frac{1}{2\kappa}\tilde{e}^{\mu[A}\tilde{e}_{\lambda}^{\;B]}\tilde{\mathcal{D}}_{\alpha}(\tilde{e}^{\alpha}_{\;A}\tilde{e}^{\lambda}_{\;B})\right)\delta\tilde{\omega}_{\sigma}^{\;\;KI}\right]\\
        = -\int d^4x\,\lambda \left[\frac{2}{3}\tilde{e}_{\mu I}\tilde{e}^{\nu}_{\;J} \left(s^{\mu}-\frac{1}{2\kappa}\tilde{e}^{\mu[A}\tilde{e}_{\lambda}^{\;B]}\tilde{\mathcal{D}}_{\sigma}(\tilde{e}^{\sigma}_{\;A}\tilde{e}^{\lambda}_{\;B})\right)\delta\tilde{\omega}_{\nu}^{\;\;IJ} + \frac{1}{2\kappa}\tilde{\mathcal{D}}_{\mu}(\tilde{e}^{[\mu}_{\;I}\tilde{e}^{\nu]}_{\;J}) \,\delta\tilde{\omega}_{\nu}^{\;\;IJ}\right.\\
        \left. + \frac{2}{3}\tilde{e}_{\mu I}\tilde{e}^{\sigma}_{\;K} \left(s^{\mu}-\frac{1}{2\kappa}\tilde{e}^{\mu[A}\tilde{e}_{\lambda}^{\;B]}\tilde{\mathcal{D}}_{\alpha}(\tilde{e}^{\alpha}_{\;A}\tilde{e}^{\lambda}_{\;B})\right)\delta\tilde{\omega}_{\sigma}^{\;\;KI}\right]
    \end{split}
\end{equation*}
It is then clear that, upon swapping the upper Lie algebra indices of $\delta\tilde{\omega}$, antisymmetry ensures that the last term cancels with the first, leaving
\begin{equation*}
    \delta\widehat{S} = -\int d^4x\,\frac{1}{2\kappa}\lambda\,\tilde{\mathcal{D}}_{\mu}(\tilde{e}^{[\mu}_{\;I}\tilde{e}^{\nu]}_{\;J})\,\delta\tilde{\omega}_{\nu}^{\;\;IJ} = 0
\end{equation*}
from which it immediately follows that
\begin{equation}\label{Eq:App1EoM1}
    \tilde{\mathcal{D}}_{\mu}(\tilde{e}^{[\mu}_{\;I}\tilde{e}^{\nu]}_{\;J})=0
\end{equation}
With this, as claimed, the second-order Herglotz Lagrangian (\ref{Eq:App1HerglotzLagrangian}) reduces to
\begin{equation}\label{Eq:App1SecondOrderL}
    L^H = \frac{1}{2\kappa} \tilde{e}^{\mu}_{\;I}\tilde{e}^{\nu}_{\;J}\tilde{R}_{\mu\nu}^{\quad IJ} - \frac{\kappa}{3} \, \tilde{e}_{\mu}^{\;K} \tilde{e}_{\nu K} s^{\mu}s^{\nu}
\end{equation}
In section (\ref{Sec:Metric}), we shifted the focus of our analysis, favouring the metric formalism. The second-order Lagrangian (\ref{Eq:App1SecondOrderL}), expressed in terms of $G_{\mu\nu}$ and $s^{\mu}$, reads
\begin{equation}\label{Eq:App1MetricL}
    L^H = \frac{1}{2\kappa}\tilde{R} - \frac{\kappa}{3}G_{\mu\nu}s^{\mu}s^{\nu}
\end{equation}
As demonstration of the consistency of our formalism, we would like to show that the full Hilbert action, when expressed in conformal variables, leads precisely to the field equations (\ref{Eq:FrictionalFieldEqns}). In doing so, we will have shown explicitly that no dynamical information is lost when eliminating the scaling degree of freedom, and that, as claimed, the conformal factor is redundant from the perspective of observable dynamics. We begin by noting that, in $d$ dimensions, two metrics $g$ and $G$ related by a conformal rescaling $g_{\mu\nu}=e^{\Phi}G_{\mu\nu}$ have Ricci scalars $R$ and $\tilde{R}$ such that
\begin{equation}\label{Eq:App1RicciRelation}
    \tilde{R} = e^{-\Phi} \left(R -(d-1)G^{\mu\nu}\tilde{\nabla}_{\mu}\tilde{\nabla}_{\nu}\Phi-\frac{(d-1)(d-2)}{4}G^{\mu\nu}\,\tilde{\nabla}_{\mu}\Phi\,\tilde{\nabla}_{\nu}\Phi\right)
\end{equation}
where, as in the main text, $\tilde{\nabla}$ denotes the connection compatible with $G_{\mu\nu}$. As discussed in detail in section (\ref{Sec:Metric}), not all metric variations are admissible. In order to correctly deduce the field equations, we should temporarily relax the assumption that $G_{\mu\nu}$ is of fixed determinant. The calculation should then be performed as described by (\ref{Eq:ConstrainedVar2}), in which we freely vary $S$, before subtracting a term proportional to the trace of this variation. Upon obtaining the final expression for the field equations, we may reinstate the condition $\textrm{det}\,G=-\,1$. With these considerations, it follows that, upon integrating by parts and assuming no contribution from boundary terms, the relevant action in four dimensions may be expressed as
\begin{equation}\label{Eq:App1EHAction}
    S=\frac{1}{2\kappa}\int d^4x \,\sqrt{-G}\,e^{\Phi}\left(\tilde{R} + \frac{3}{2}G^{\mu\nu}\partial_{\mu}\Phi\,\partial_{\nu}\Phi\right)
\end{equation}
Variations of this action with respect to $G^{\mu\nu}$ yield
\begin{equation*}
    \delta S = \frac{1}{2\kappa}\int d^4x \,\sqrt{-G}e^{\Phi}\left(-\frac{1}{2}\left(\tilde{R}+\frac{3}{2}G^{\alpha\beta}\partial_{\alpha}\Phi\,\partial_{\beta}\Phi \right) G_{\mu\nu}\,\delta G^{\mu\nu} + \delta G^{\mu\nu} \tilde{R}_{\mu\nu} + G^{\mu\nu}\,\delta\tilde{R}_{\mu\nu} + \frac{3}{2}\delta G^{\mu\nu}\,\partial_{\mu}\Phi\,\partial_{\nu}\Phi\right)
\end{equation*}
The behaviour of $\tilde{R}_{\mu\nu}$ under a change $\delta G^{\mu\nu}$ is a standard result \cite{wald2010general}
\begin{equation}
    \delta\tilde{R}_{\mu\nu} = \tilde{\nabla}_{\rho}\delta\tilde{\Gamma}_{\mu\nu}^{\rho} - \tilde{\nabla}_{\mu}\delta\tilde{\Gamma}^{\rho}_{\rho\nu}
\end{equation}
We thus need only focus on the manipulation of the following term
\begin{equation*}
    \delta S \,\supset\, \frac{1}{2\kappa}\int d^4x\,\sqrt{-G}e^{\Phi} \,\tilde{\nabla}_{\rho}\left(G^{\mu\nu}\,\delta\tilde{\Gamma}^{\rho}_{\mu\nu} - G^{\rho\nu}\,\delta\tilde{\Gamma}^{\lambda}_{\lambda\nu}\right):=I_1-I_2
\end{equation*}
The variation of the connection coefficients may be expressed as
\begin{equation}
    \delta \tilde{\Gamma}_{\mu\nu}^{\rho} = \frac{1}{2}G^{\alpha\lambda}\left(\tilde{\nabla}_{\nu}\delta G_{\mu\alpha} + \tilde{\nabla}_{\mu} \delta G_{\nu\alpha} - \tilde{\nabla}_{\alpha} \delta G_{\mu\nu} \right)
\end{equation}
We may then write the integral $I_1$ as follows (throughout, we do not retain total divergences)
\begin{align*}
    I_1 &:= \frac{1}{2\kappa}\int d^4x\,\sqrt{-G}e^{\Phi} \,\tilde{\nabla}_{\rho} G^{\mu\nu}\,\delta\tilde{\Gamma}^{\rho}_{\mu\nu}\\
    &= -\,\frac{1}{4\kappa}\int d^4x \,\sqrt{-G}e^{\Phi}\,(\partial_{\rho}\Phi) G^{\mu\nu} G^{\alpha\lambda}\left(\tilde{\nabla}_{\nu}\delta G_{\mu\alpha} + \tilde{\nabla}_{\mu} \delta G_{\nu\alpha} - \tilde{\nabla}_{\alpha} \delta G_{\mu\nu} \right)\\
    &= -\,\frac{1}{4\kappa}\int d^4x \,\sqrt{-G}e^{\Phi}\,(\partial_{\rho}\Phi)\left(2\,\tilde{\nabla}_{\nu}\left(G^{\mu\nu} G^{\alpha\lambda} \,\delta G_{\mu\alpha}\right) - \tilde{\nabla}_{\alpha}\left(G^{\mu\nu} G^{\alpha\lambda} \,\delta G_{\mu\nu}\right) \right)\\
    &= \frac{1}{4\kappa}\int d^4x \,\sqrt{-G}e^{\Phi}\biggr[2\left(\partial_{\nu}\Phi\,\partial_{\rho}\Phi + \tilde{\nabla}_{\nu} (\partial_{\rho}\Phi)\right) G^{\mu\nu} G^{\alpha\lambda} \,\delta G_{\mu\alpha} - \left(\partial_{\alpha}\Phi\,\partial_{\rho}\Phi + \tilde{\nabla}_{\alpha} (\partial_{\rho}\Phi)\right)G^{\mu\nu} G^{\alpha\lambda} \,\delta G_{\mu\nu}\biggr]\\
    &=-\, \frac{1}{4\kappa}\int d^4x \,\sqrt{-G}e^{\Phi}\biggr[2\left(\partial_{\mu}\Phi\,\partial_{\nu}\Phi + \tilde{\nabla}_{\mu} (\partial_{\nu}\Phi)\right) - G^{\alpha\beta}\left( \partial_{\alpha}\Phi\,\partial_{\beta}\Phi + \tilde{\nabla}_{\alpha} (\partial_{\beta}\Phi)\right)G_{\mu\nu}\biggr]\,\delta G^{\mu\nu}
\end{align*}
In passing to the final line, we have used that $\delta G_{\alpha\beta}=-\, G_{\alpha\mu} G_{\beta\nu}\,\delta G^{\mu\nu}$. A series of entirely analogous manipulations are carried out, yielding the following form of $I_2$
\begin{equation}\label{Eq:I2}
    I_2 = -\,\frac{1}{4\kappa}\int d^4x \,\sqrt{-G}e^{\Phi} G^{\alpha\beta}\left( \partial_{\alpha}\Phi\,\partial_{\beta}\Phi + \tilde{\nabla}_{\alpha} (\partial_{\beta}\Phi)\right)G_{\mu\nu}\,\delta G^{\mu\nu}
\end{equation}
Consequently, we find that the term in $\delta S$ containing $\delta\tilde{R}_{\mu\nu}$ may be expressed as
\begin{equation}
    -\, \frac{1}{2\kappa}\int d^4x \,\sqrt{-G}e^{\Phi}\biggr[\left(\partial_{\mu}\Phi\,\partial_{\nu}\Phi + \tilde{\nabla}_{\mu} (\partial_{\nu}\Phi)\right) - G^{\alpha\beta}\left( \partial_{\alpha}\Phi\,\partial_{\beta}\Phi + \tilde{\nabla}_{\alpha} (\partial_{\beta}\Phi)\right)G_{\mu\nu}\biggr]\,\delta G^{\mu\nu}
\end{equation}
With this, the full (unconstrained) variation of the action with respect to $G^{\mu\nu}$ reads
\begin{equation}
    \begin{split}
        \delta S = \frac{1}{2\kappa}\int d^4x \,\sqrt{-G}e^{\Phi}\biggr[ \tilde{R}_{\mu\nu} -\frac{1}{2}\left(\tilde{R} - \frac{1}{2}G^{\alpha\beta} \partial_{\alpha}\Phi\,\partial_{\beta}\Phi - 2\, G^{\alpha\beta}\tilde{\nabla}_{\alpha}(\partial_{\beta}\Phi)\right)G_{\mu\nu} \\
        + \, \frac{1}{2}\partial_{\mu}\Phi\,\partial_{\nu}\Phi - \frac{1}{2}\left(\tilde{\nabla}_{\mu}(\partial_{\nu}\Phi) + \tilde{\nabla}_{\nu}(\partial_{\mu}\Phi)\right)\biggr]\,\delta G^{\mu\nu}
    \end{split}
\end{equation}
Since we would like the final field equations to be symmetric in the lower indices, we have symmetrised the final term. The two steps that remain are application of (\ref{Eq:ConstrainedVar2}), and elimination of $\partial_{\mu}\Phi$ in favour of the action density. For the former, we reproduce the relevant expression for the admissible variations
\begin{equation}\label{Eq:App1ConstrainedVar}
    \frac{\delta S}{\delta^A G^{\mu\nu}} = \frac{\delta S}{\delta G^{\mu\nu}} - \frac{1}{4}G^{\mu\nu}G^{\alpha\beta}\frac{\delta S}{\delta G^{\alpha\beta}}
\end{equation}
From this, we find that the equation of motion in conformal variables reads
\begin{equation*}
    \tilde{R}_{\mu\nu} -\frac{1}{4}\left(\tilde{R} + \frac{1}{2}G^{\alpha\beta} \partial_{\alpha}\Phi\,\partial_{\beta}\Phi -  G^{\alpha\beta}\tilde{\nabla}_{\alpha}(\partial_{\beta}\Phi)\right)G_{\mu\nu} + \frac{1}{2}\partial_{\mu}\Phi\,\partial_{\nu}\Phi -\frac{1}{2}\left(\tilde{\nabla}_{\mu}(\partial_{\nu}\Phi) + \tilde{\nabla}_{\nu}(\partial_{\mu}\Phi)\right)=0
\end{equation*}
Finally, the velocities $\partial_{\mu}\Phi$ of the conformal factor are replaced with the appropriate action-dependent expressions, leading to
\begin{equation}
    \begin{split}
        \tilde{R}_{\mu\nu} - \frac{1}{4}\left(\tilde{R}+\frac{2\kappa^2}{9}G_{\alpha\beta}s^{\alpha}s^{\beta} - \frac{2\kappa}{3}\partial_{\alpha}s^{\alpha}\right)G_{\mu\nu} + \frac{2\kappa^2}{9}G_{\mu\alpha}G_{\nu\beta}s^{\alpha}s^{\beta} \\
        -\frac{\kappa}{3}\left(G_{\mu\lambda}\tilde{\nabla}_{\nu}s^{\lambda} + G_{\lambda\nu}\tilde{\nabla}_{\mu}s^{\lambda}\right)=0
    \end{split}
\end{equation}
precisely as in (\ref{Eq:FrictionalFieldEqns}).
\bibliographystyle{unsrt}
\bibliography{Refs}
\end{document}